\documentclass[12pt, a4paper]{article}

\usepackage[margin=1in]{geometry}

\usepackage[utf8]{inputenc}
\usepackage[T1]{fontenc}
\usepackage{lmodern}

\usepackage{amsmath, amssymb}

\usepackage{graphicx}
\usepackage{tikz}
\usetikzlibrary{positioning, arrows.meta, fit, backgrounds, shapes.geometric, calc, decorations.pathreplacing}
\usepackage{pgfplots}
\pgfplotsset{compat=1.18}
\usepackage{tabularx}

\usepackage{booktabs}
\usepackage{array}
\usepackage{multirow}
\usepackage{ragged2e}
\newcolumntype{Y}{>{\raggedright\arraybackslash}X}

\usepackage{microtype}
\usepackage[hyphens]{url}
\usepackage{xurl}

\usepackage[numbers,square]{natbib}
\usepackage[hidelinks, hypertexnames=false]{hyperref}

\usepackage{enumitem}
\usepackage{xcolor}
\usepackage{float}
\usepackage{authblk}
\usepackage{indentfirst}
\makeatletter
\renewenvironment{abstract}{%
  \par\vspace{-0.5em}%
  \begin{center}
    \bfseries \abstractname
  \end{center}
  \begin{list}{}{\leftmargin=1.5em \rightmargin=1.5em}
  \item\relax
  \footnotesize
  \setlength{\parindent}{0pt}
  \noindent
  \ignorespaces
}{%
  \end{list}
}
\makeatother
\AtBeginEnvironment{thebibliography}{\sloppy}

\title{\textbf{ASTELD: A Six-Axis Classification Framework for Autonomous AI Agents --- Design, Evaluation, and an OpenClaw Case Study}}

\author[1,9]{Siyuan Li}
\author[1]{Peng Shu}
\author[2]{Churan Yu}
\author[3]{Peilong Wang}
\author[1]{Ruidong Zhang}
\author[4]{Bowen Guo}
\author[1]{Xinliang Li}
\author[5]{Ruiyu Yan}
\author[6]{Arif Hassan Zidan}
\author[1]{Yi Pan}
\author[1]{Wei Ruan}
\author[1]{Lifeng Chen}
\author[1]{Junhao Chen}
\author[1]{Zhaojun Ding}
\author[1]{Yiwei Li}
\author[1]{Zhengliang Liu}
\author[7]{Haixing Dai}
\author[8]{Lin Zhao}
\author[4]{Yu Bao}
\author[9]{Xiang Li}
\author[6]{Wei Zhang}
\author[1]{Tianming Liu\thanks{Correspondence: \href{mailto:tliu@uga.edu}{tliu@uga.edu}}}

\affil[1]{School of Computing, University of Georgia, Athens, GA, USA}
\affil[2]{College of Engineering, University of Georgia, Athens, GA, USA}
\affil[3]{Department of Radiation Oncology, City of Hope National Medical Center, Duarte, CA, USA}
\affil[4]{Department of Graduate Psychology, James Madison University, VA, USA}
\affil[5]{Tandon School of Engineering, New York University, New York, NY, USA}
\affil[6]{School of Computer and Cyber Sciences, Augusta University, Augusta, GA, USA}
\affil[7]{Meta, Menlo Park, CA, USA}
\affil[8]{Department of Biomedical Engineering, New Jersey Institute of Technology, Newark, NJ, USA}
\affil[9]{Department of Radiology, Massachusetts General Hospital and Harvard Medical School, Boston, MA, USA}

\date{}

\begin{document}
\maketitle
\vspace{-1.2em}

\begin{abstract}
Autonomous AI agent platforms differ substantially in architecture, security, tool integration, execution, autonomy, and deployment, yet the field lacks a common classification scheme for comparing these design choices. We propose ASTELD, an operational six-axis classification framework for autonomous AI agents: Architecture pattern, Security posture, Tool integration model, Execution paradigm, Level of autonomy and human control, and Deployment topology. ASTELD is constructed by synthesizing prior agent taxonomies with observable platform properties and explicit category-assignment rules. We evaluate its discriminative and explanatory utility by mapping eight representative frameworks and by using OpenClaw as an in-depth case study. The resulting profiles separate all eight platforms under their dominant configurations and reveal three cross-platform patterns: a security--accessibility diagonal, strong execution--architecture coupling, and capability convergence with persistent architectural differentiation. We further classify 50+ OpenClaw derivatives and find that innovation concentrates on the Security, Execution, and Deployment axes, indicating that ASTELD can explain where ecosystem fragmentation occurs. The OpenClaw case study also supplies a six-category vulnerability taxonomy, evidence from five institutional assessments, and adoption and governance analyses that connect platform coordinates to observed risks. These results position ASTELD as a reproducible method for comparing agent platforms, identifying unoccupied design regions, guiding framework selection, and organizing future empirical research. The analysis also exposes a consequential empty region: none of the evaluated systems combines local-first deployment with enterprise-grade security.
\end{abstract}

\begin{list}{}{\leftmargin=1.5em \rightmargin=1.5em}
\item\relax
\footnotesize
\setlength{\parindent}{0pt}
\end{list}

\newpage
\tableofcontents
\newpage

\section{Introduction}
\label{sec:intro}

The emergence of autonomous artificial intelligence (AI) agents represents a paradigmatic shift in how humans interact with AI. Unlike chatbots that generate text in response to prompts, autonomous agents decompose goals into subtasks, invoke tools, access file systems, and communicate with external services---all with minimal human oversight. This transition from \emph{generation} to \emph{action} has profound implications for software engineering, security, and governance.

These implications are no longer theoretical. In the span of a few months, a single open-source project, OpenClaw, has made them concrete. Released as a weekend side project in November~2025 under the name \textit{Clawdbot}, OpenClaw accumulated 60,000 GitHub stars within four days of going viral in late January~2026, surpassed React as the most-starred software project on GitHub by March~3, 2026 with 250,829~stars, and reached 360,000+ stars on GitHub by late April~2026~\cite{openclawvps2026statistics}\cite{aftab2026react_record}\cite{openclawblog2026react_milestone}\cite{github_openclaw}. In parallel, public trackers documented 100+ tracked advisories and 10+ published CVEs by April~2026, while broader vendor summaries reported larger tallies of tracked vulnerability records under different aggregation rules~\cite{jgamblin2026openclawcves}\cite{blink2026security}; OpenClaw also suffered a coordinated supply chain attack, \textit{ClawHavoc}, that ultimately led to the identification of 1,184 malicious skills in its marketplace~\cite{cyberpress2026clawhavoc}\cite{koi2026clawhavoc}. This is the paradox: the most-starred repository on GitHub is also a case study in security debt. The tensions it reveals are not unique to one project; they are built into the agent paradigm itself.

Three such tensions recur throughout this analysis. First, the \textit{accessibility--security trade-off}: the design choices that made OpenClaw accessible to non-technical users (a messaging-platform interface, a monolithic daemon with system-level access, a permissionless skill marketplace) are precisely the choices that created its most serious vulnerabilities. Second, the \textit{open-source speed paradox}: rapid community-driven development outpaced the project's capacity to establish security governance, resulting in a 4-day burst in which 9 new security-tracker entries were logged between March~18 and March~21, 2026~\cite{jgamblin2026openclawcves}. Third, the \textit{impossibility of architectural universality}: no single agent architecture can simultaneously optimize for accessibility, security, expressiveness, and enterprise readiness---a finding we demonstrate through systematic cross-framework comparison.

Existing studies describe individual agent platforms, compare selected frameworks, or survey security and ecosystem issues~\cite{suwansathit2026openclaw_vulnerabilities}\cite{shan2026claw_grip}\cite{preprints2026openclaw_survey}\cite{manik2026moltbook}\cite{openclaw2026agent_runtime}. What remains missing is an operational classification method that converts heterogeneous platform evidence into comparable coordinates. Without explicit axes and assignment rules, comparisons remain narrative, framework selection is difficult to reproduce, and ecosystem changes cannot be related systematically to the design constraints of a base platform.

We formulate the problem as follows: given publicly observable evidence about an autonomous-agent platform, assign a compact profile that captures its dominant design while preserving meaningful configuration ranges. The classification should (i) distinguish platforms with different design commitments, (ii) reveal cross-axis regularities and unoccupied design regions, and (iii) help explain how derivative systems modify the constraints of a base platform. OpenClaw provides a high-information case for this task because its architecture, security incidents, rapid adoption, and unusually large derivative ecosystem expose the consequences of platform design choices at scale.

To address this problem, we propose and evaluate ASTELD, a six-axis classification framework covering Architecture, Security, Tool integration, Execution, Level of autonomy and human control, and Deployment. The paper answers three classification questions: whether observable platform properties can be encoded consistently along these axes; whether the resulting profiles discriminate among representative frameworks and expose structural patterns; and whether axis-level constraints help explain the direction of derivative innovation.

Specifically, we make five contributions:

\begin{enumerate}

\item\textbf{ASTELD classification method.} We introduce a six-axis taxonomy with explicit, observable category definitions for classifying autonomous AI agent platforms. The method supports dominant-value assignments, range-valued profiles, and transition notation when common configurations span categories.

\item\textbf{Empirical framework evaluation.} We operationalize ASTELD on eight representative platforms and evaluate its discriminative and analytical utility through profile separation, cross-axis pattern analysis, and identification of unoccupied design regions.

\item\textbf{OpenClaw case evidence.} We construct a multi-dimensional evidence base covering OpenClaw's architecture, security, adoption, governance, and ecosystem position. This includes a six-category vulnerability taxonomy synthesized from CVE databases, five institutional assessments (Microsoft~\cite{microsoft2026running_openclaw}, CertiK~\cite{certik2026openclaw_report}, CrowdStrike~\cite{zaitsev2026security_teams}, Trend Micro~\cite{tucci2026cisos}, and Koi Security~\cite{koi2026clawhavoc}), and academic research.

\item\textbf{Derivative-based explanatory analysis.} We classify 50+ OpenClaw derivatives across three tiers and test whether observed modifications concentrate on the ASTELD axes that are most constrained in the base platform. The results show concentration on Security, Execution, and Deployment.

\item\textbf{Research and selection implications.} We derive a practitioner selection matrix, identify the local-first plus enterprise-security design gap, and formulate eight research questions spanning architectural, security, ecosystem, and sociotechnical domains.

\end{enumerate}

The remainder of this paper is organized as follows. Section~\ref{sec:background} establishes the intellectual context by tracing the evolution from chatbots to autonomous agents. Section~\ref{sec:origins} introduces OpenClaw's origins, evolution, and governance transitions. Section~\ref{sec:architecture} provides a technical deep dive into its architecture. Section~\ref{sec:security} presents our security and privacy analysis, including the six-category vulnerability taxonomy and attack chain composition. Section~\ref{sec:ecosystem} positions OpenClaw within the broader AI agent ecosystem through the ASTELD classification framework. Section~\ref{sec:adoption} examines adoption dynamics, social impact, and the derivative ecosystem. Section~\ref{sec:rq} formulates open research questions and future directions. Finally, Section~\ref{sec:conclusion} concludes the work.

\section{{Background: From Chatbots to Autonomous AI Agents}}
\label{sec:background}

\subsection{LLM-Based Applications}
\label{sec:bg-generations}

The development of LLM-based applications can be understood as a progression from standalone text generation, to retrieval- and workflow-enhanced systems, and more recently to autonomous agentic architectures.

\textbf{Text Generation Chatbots.} The release of GPT-3~\cite{brown2020gpt3} established the paradigm of prompt-in, text-out interaction. Users provided natural language instructions; models returned generated text. Applications included summarization, translation, and question answering. The model had no memory, no tool access, and no ability to take actions in the world.

\textbf{Retrieval-Augmented Systems.} Retrieval-Augmented Generation (RAG) extended LLMs by grounding their outputs in external knowledge bases. Frameworks such as LangChain (released October~2022)~\cite{langchain2025langgraph} provided composable abstractions for chaining LLM calls with document retrieval, enabling applications like domain-specific chatbots and enterprise knowledge assistants. The model could now reference external data, but still could not act on the world.

\textbf{Autonomous AI Agents.} The agent paradigm represents a qualitative shift: LLMs are no longer endpoints but controllers. An autonomous agent observes its environment, formulates plans, invokes tools (APIs, shell commands, file operations), and iterates until a goal is achieved. AutoGPT, released in March~2023~\cite{github_autogpt}, was the first viral demonstration of this paradigm. The subsequent emergence of multi-agent frameworks---AutoGen~\cite{autogen2023framework} for multi-agent conversation, CrewAI~\cite{agenttimes2026crewai} for role-based collaboration, LangGraph~\cite{langchain2025langgraph} for stateful graph execution---established autonomous agents as a recognized architectural pattern.

OpenClaw is significant as it represents a further step in the operationalization of autonomous agents. By combining an LLM controller with system-level OS access, a messaging-platform UI (WhatsApp, Telegram, Discord), and an open plugin marketplace, OpenClaw made autonomous AI agents accessible to non-technical users for the first time~\cite{milvus2026openclaw_guide}\cite{kdnuggets2026openclaw}. This democratization is simultaneously its greatest achievement and its most serious liability.

\subsection{Definition of Autonomous AI Agents}
\label{sec:bg-definition}

We adopt the following operational definition, drawing on recent surveys of agentic AI and autonomous agents~\cite{v2026agentic_ai}\cite{xu2026ai_agent_systems}\cite{wang2023autonomous_agents_survey}\cite{sapkota2025agentic_ai}:

\begin{quote}
An \textbf{autonomous AI agent} is a software system in which a large language model serves as a cognitive controller that (1)~maintains persistent state across interactions, (2)~formulates multi-step plans to achieve user-specified goals, (3)~invokes external tools and services to execute those plans, and (4)~operates with a degree of autonomy that permits action without per-step human approval.
\end{quote}

This definition distinguishes agents from chatbots (which lack tool invocation and persistent state), from RAG systems (which lack planning and action), and from traditional automations (which are not LLM or AI-based). The Interface EU classification~\cite{interfaceeu2024autonomy} further stratifies autonomy into five levels, from L1 (tool mode, every action requires instruction) through L5 (full autonomous operation, no human oversight). As we show in Section~\ref{sec:architecture}, OpenClaw's design enables autonomous operations up to L4 (agent decomposes and executes complex goals end-to-end)~\cite{microsoft2026running_openclaw}.

\subsection{The Security Implications of the Agent Paradigm}
\label{sec:bg-security}

The transition from chatbots to agents introduces a fundamentally broader security surface. A useful framing, sometimes described as the ``lethal trifecta,'' is the combination of (1) access to private data on the user's device, (2) exposure to untrusted external content from external sources, and (3) the ability to communicate with or act upon external services. When all three conditions hold, prompt injection becomes not merely a content moderation problem but an authorization bypass---a malicious instruction embedded in retrieved content can cause the agent to exfiltrate data, install software, or modify system configurations~\cite{microsoft2026running_openclaw}\cite{penligent2026redteam}\cite{arnav2026prompt_injection}.

This threat model is not hypothetical. Public reporting on OpenClaw in early 2026 documented a rapidly growing stream of vulnerabilities and abuse cases, including CVE-2026-25253 (ClawJacked: cross-site WebSocket hijacking, CVSS~8.8)~\cite{oasis2026clawjacked} and CVE-2026-32922 (privilege escalation)~\cite{armo2026cve32922}. The ClawHavoc campaign demonstrated that supply chain attacks against agent plugin marketplaces can achieve unprecedented scale: Koi Security's initial audit identified 341 malicious skills among 2,857 on ClawHub (approximately 12\%), and subsequent reporting later placed the total at 1,184 as the investigation widened~\cite{cyberpress2026clawhavoc}\cite{koi2026clawhavoc}. These incidents are detailed in Section~\ref{sec:security}.

OpenClaw emerged at the inflection point of this transition---and its trajectory reveals both the promise and the peril of the agent paradigm.

\section{OpenClaw: Origins and Evolution}
\label{sec:origins}

\subsection{From Clawdbot to OpenClaw}

OpenClaw began its life as \textbf{Clawdbot}, a weekend side project created by
Austrian developer Peter Steinberger in November~2025~\cite{wikipedia2026openclaw}. Steinberger,
previously known as the founder of PSPDFKit (a cross-platform PDF SDK company),
built Clawdbot as a personal experiment: a daemon process that connected a large
language model to the local operating system via messaging platforms---WhatsApp,
Telegram, and later Discord and Slack. The original vision was deceptively
simple: enable users to interact with their computers through natural language
messages sent from their phones~\cite{steinberger2026openclaw}\cite{openclawblog2026foundation}.

In its earliest form, Clawdbot ran as a local process on macOS, accepting
instructions via WhatsApp and executing them through shell commands, file
operations, and API calls. There was no plugin system, no marketplace, and no
multi-user support. The project attracted modest attention in its first two
months---respectable but unremarkable by open-source standards~\cite{openclawvps2026statistics}.

The transformation from niche tool to global phenomenon began on
January~29, 2026, when the project was rebranded to \textbf{OpenClaw} after a
brief interim rename to \textbf{Moltbot} on January~27 following trademark
concerns raised by Anthropic. The OpenClaw rebrand coincided with a restructured
architecture that included a skill system, a nascent extension marketplace
(ClawHub), and support for multiple LLM backends~\cite{openclawvps2026statistics}. The timing coincided with a wave of social media
attention: within 48~hours, OpenClaw was accumulating stars at a rate of 710 per
hour---17,084 per day---a velocity without precedent in GitHub's
history~\cite{openclawvps2026statistics}\cite{aftab2026react_record}. By February~2, 2026, within four days of the rebrand, the project had accumulated over 60,000 new stars~\cite{openclawvps2026statistics}.

\subsection{Growth Trajectory and Milestones}

The growth of OpenClaw can be divided into four phases, each with distinct dynamics (see Figure~\ref{fig:growth}):

\textbf{Phase~1: Viral Ignition (January~29 -- February~2, 2026).} The initial burst was driven by technology media coverage and social media amplification. The project's appeal was immediate: unlike existing AI agent frameworks that required programming knowledge, OpenClaw could be installed by non-technical users and accessed through familiar messaging interfaces. This four-day window produced 60,000~stars at an average of 15,000 per day~\cite{openclawvps2026statistics}\cite{aftab2026react_record}.

\textbf{Phase~2: Sustained Acceleration (February~2 -- March~3, 2026).} The viral wave did not follow the typical pattern of rapid decay. Instead, growth sustained at approximately 6,600~stars per day, driven by tutorial content, derivative projects, and enterprise experimentation. By February~24, OpenClaw had crossed 224,000~stars and surpassed the Linux kernel (218,000~stars accumulated over 30~years), becoming one of the most-starred repositories in GitHub history~\cite{aftab2026react_record}\cite{starhistory2026openclaw}. The velocity comparison is instructive: Linux accumulated 100,000~stars in approximately 30~years (20 per day average); React reached 100,000 in approximately 10~years (68 per day); OpenClaw reached 100,000 in 12~days---a velocity 123~times that of React and 419~times that of Linux~\cite{ossinsight2026forks_wave}\cite{starhistory2026openclaw}.

\textbf{Phase~3: China Wave (March~3 -- March~24, 2026).} After a brief deceleration, OpenClaw experienced a pronounced March reacceleration driven by adoption in China. Traffic from Chinese IP addresses surged by 1,436\% month-over-month, propelled by what Chinese media termed the ``lobster craze'' (l\'{o}ngxi\={a} r\`{e}ch\'{a}o, a wordplay on ``Claw'')~\cite{gradually2026statistics}\cite{starhistory2026openclaw}. During this phase, OpenClaw surpassed React (243,438~stars accumulated over 10~years) on March~3, 2026, at 250,829~stars, and was officially recognized as the most-starred software project on GitHub on the same day~\cite{aftab2026react_record}. Growth averaged approximately 4,000~stars per day during this phase, reaching 335,000 by March~24~\cite{openclawvps2026statistics}\cite{gradually2026statistics}.

\textbf{Phase~4: Stabilization (March~24 -- present).} Growth decelerated to under 1,000~stars per day, reaching 360,000+ stars on GitHub by late April~2026~\cite{gradually2026statistics}\cite{starhistory2026openclaw}\cite{github_openclaw}. This phase coincided with increasing media coverage of security vulnerabilities and the disclosure of the ClawHavoc supply chain campaign.

The growth trajectory is unusual: rather than following a simple post-viral decay curve, OpenClaw exhibited an abrupt late-January inflection, sustained acceleration through mid-February, and a later March reacceleration associated with Chinese adoption. This pattern suggests that the project's growth was not a single viral event but a compounding phenomenon in which each adoption wave created conditions for the next~\cite{ossinsight2026forks_wave}.

\begin{table}[tbp]
\centering
\caption{Timeline of OpenClaw milestones.}
\label{tab:timeline}
\footnotesize
\setlength{\tabcolsep}{4pt}
\begin{tabularx}{\textwidth}{@{}>{\RaggedRight\arraybackslash}p{1.9cm} >{\RaggedRight\arraybackslash}p{2.0cm} Y >{\RaggedRight\arraybackslash}p{2.85cm}@{}}
\toprule
\textbf{Date} & \textbf{Event} & \textbf{Description} & \textbf{Ref.} \\
\midrule
2025-11-24 & Clawdbot launched       & Peter Steinberger releases personal AI agent as weekend project; macOS-only, WhatsApp interface, Claude backend & \cite{wikipedia2026openclaw} \\
2026-01-29 & Rebranded to OpenClaw   & Final rebrand after the brief Moltbot interim name; model-agnostic architecture, ClawHub skill marketplace, and multi-platform messaging support debut together & \cite{openclawvps2026statistics} \\
2026-01-30 & Peak viral day          & 710 stars/hour (17{,}084/day) reported within the first 48 hours after the OpenClaw rebrand & \cite{openclawvps2026statistics}\cite{aftab2026react_record} \\
2026-02-02 & 60,000 stars & 60K reached within four days post-rebrand; Phase 1 (Viral Ignition) concludes & \cite{openclawvps2026statistics} \\
2026-02-07 & ClawHavoc disclosed     & Koi Security identifies 341 malicious skills among 2{,}857 on ClawHub ($\sim$12\%); subsequent reporting later places the total at 1{,}184 & \cite{cyberpress2026clawhavoc}\cite{koi2026clawhavoc} \\
2026-02-12 & 190{,}000 stars         & 190K in 14 days; NanoClaw, ZeroClaw, PicoClaw, Nanobot forks emerge within a 14-day window & \cite{openclawvps2026statistics}\cite{ossinsight2026forks_wave} \\
2026-02-15 & Steinberger joins OpenAI & Creator hired by OpenAI; foundation governance announced for OpenClaw & \cite{openclawblog2026foundation} \\
2026-02-24 & Surpassed Linux kernel  & Crosses 224K+ stars and exceeds Linux's 218K stars (accumulated over 30 years) & \cite{aftab2026react_record}\cite{starhistory2026openclaw} \\
2026-03-03 & Surpassed React         & Exceeds React's 243{,}438 stars (accumulated over 10 years) & \cite{aftab2026react_record} \\
2026-03-03 & Most-starred repository & 250{,}829 stars; officially the most-starred software project on GitHub & \cite{aftab2026react_record} \\
2026-03-18--21 & Peak disclosure burst & 9 new security-tracker entries logged in 4 days, including critical authentication bypasses & \cite{jgamblin2026openclawcves} \\
2026-03-24 & China wave peak         & 335K stars; Chinese traffic +1{,}436\% MoM; ``lobster craze'' phenomenon & \cite{gradually2026statistics}\cite{starhistory2026openclaw} \\
Late Apr.~2026 & Current                 & 360K+ stars on GitHub; 3M+ MAU; 40K+ ClawHub skills; 1,800+ repo contributors & \cite{github_openclaw}\cite{gradually2026statistics}\cite{openclawvps2026statistics} \\
\bottomrule
\end{tabularx}
\end{table}

\subsection{Governance Transitions}
OpenClaw's governance has undergone three transitions in rapid succession:

\textbf{Solo maintainer (November~2025 -- January~2026).} In its Clawdbot
phase, Steinberger was the sole maintainer, making all architectural and
release decisions.

\textbf{Community-governed open source (January -- February~2026).} The
rebrand to OpenClaw coincided with the opening of contributions. The
contributor base grew from 1 to 1,800+~contributors by late April~2026~\cite{ossinsight2026forks_wave}\cite{github_openclaw}.
However, the project lacked formal governance structures---no steering
committee, no security response team, and no code review requirements for
skill submissions to ClawHub.

\textbf{Foundation transition (February~2026 -- present).} Steinberger's hire
by OpenAI in February~2026~\cite{openclawblog2026foundation} created uncertainty about the project's
independence. In response, a foundation structure was announced to steward the
project. Notably, this governance transition coincided with OpenClaw's peak
security-disclosure burst: 9 entries were added in a 4-day window (March~18--21,
2026)~\cite{jgamblin2026openclawcves}. Whether the transition affected security accountability
remains unclear; the disclosure spike may reflect increased researcher scrutiny
of a high-profile project as much as any governance gap.

\subsection{Naming and Identity}
The project's naming history reflects its evolving identity. The original name
``Clawdbot'' was a portmanteau of ``Claude'' (the Anthropic LLM that served as
its initial backend) and ``bot.'' Following trademark concerns raised by
Anthropic, the project was briefly renamed ``Moltbot'' on January~27, 2026,
before settling on ``OpenClaw'' on January~29, 2026~\cite{openclawvps2026statistics}. The final rebrand
signaled two shifts: the move to model-agnosticism (dropping the Claude-specific
reference) and the adoption of ``Open'' to emphasize its open-source nature. The
community has subsequently adopted the claw/lobster motif as a cultural identity
marker, with derivative projects adopting names like NanoClaw, ZeroClaw,
PicoClaw, and Dr.~Claw~\cite{ossinsight2026forks_wave}.

\subsection{Summary}

OpenClaw's evolution from a weekend experiment to the most-starred software
project in GitHub history occurred in under five months. This trajectory was
enabled by three factors: (1)~the accessibility of its messaging-platform
interface, which lowered the barrier to autonomous AI from developers to general
users; (2)~the permissiveness of its architecture, which granted the LLM
controller system-level OS access; and (3)~the openness of its extension
ecosystem, which allowed anyone to publish skills with minimal vetting. As we
show in the following sections, these same three factors created the conditions
for OpenClaw's most serious security vulnerabilities.

\section{System Architecture}
\label{sec:architecture}

This section provides a technical analysis of OpenClaw's architecture as of version 2026.3 (March~2026), drawing on official documentation~\cite{openclaw2026agent_runtime}, source code analysis, and third-party security assessments~\cite{microsoft2026running_openclaw}\cite{certik2026openclaw_report}\cite{zaitsev2026security_teams}.

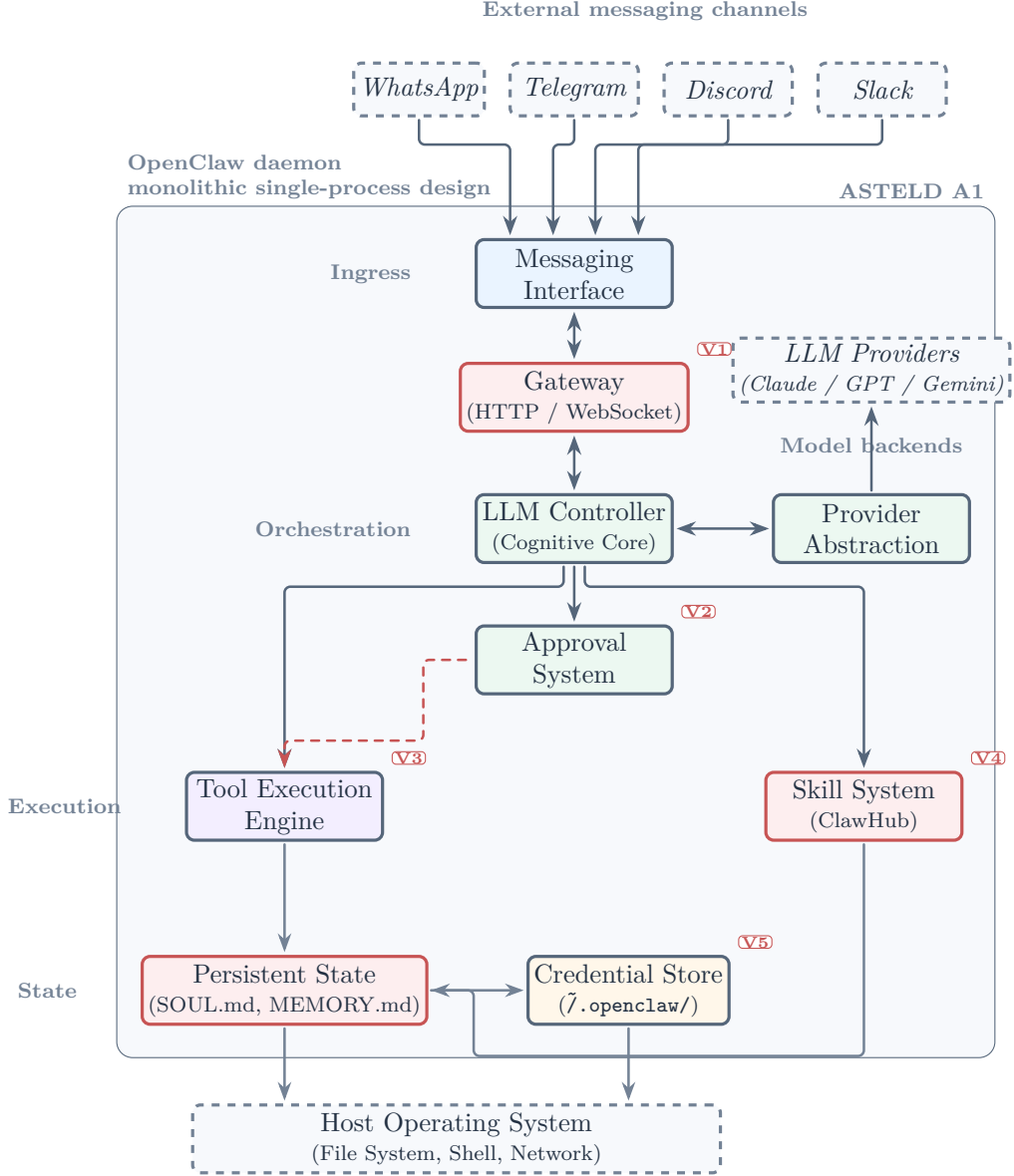
\begin{figure}[tbp]
\centering

\definecolor{ink}{HTML}{223042}
\definecolor{line}{HTML}{55667A}
\definecolor{muted}{HTML}{738295}
\definecolor{panel}{HTML}{F7F9FC}
\definecolor{iface}{HTML}{EAF4FF}
\definecolor{control}{HTML}{ECF8F1}
\definecolor{exec}{HTML}{F3EEFF}
\definecolor{state}{HTML}{FFF7EA}
\definecolor{riskfill}{HTML}{FDEEEF}
\definecolor{risk}{HTML}{C75454}

\begin{tikzpicture}[
    font=\footnotesize,
    >={Stealth[length=2.6mm,width=1.9mm]},
    node distance=8mm and 10mm,
    box/.style={
        rectangle, rounded corners=3pt, draw=line, very thick,
        minimum width=26mm, minimum height=9mm,
        align=center, fill=white, text=ink, inner sep=2.2pt
    },
    ext/.style={
        box, draw=muted, dashed, fill=panel,
        minimum width=17mm, minimum height=7mm,
        font=\footnotesize\itshape
    },
    ifacebox/.style={box, fill=iface},
    controlbox/.style={box, fill=control},
    execbox/.style={box, fill=exec},
    statebox/.style={box, fill=state},
    riskbox/.style={box, fill=riskfill, draw=risk},
    rtag/.style={
        font=\tiny\bfseries,
        text=risk,
        fill=white,
        draw=risk,
        rounded corners=1.5pt,
        inner xsep=1.1pt,
        inner ysep=0.3pt
    },
    layer/.style={font=\scriptsize\bfseries, text=muted},
    pipe/.style={
        draw=line,
        line width=0.95pt,
        line cap=round,
        line join=round,
        rounded corners=2.2pt
    },
    flow/.style={
        pipe,
        -{Stealth[length=2.6mm,width=1.9mm]},
        shorten <=1pt,
        shorten >=1pt
    },
    bflow/.style={
        pipe,
        {Stealth[length=2.6mm,width=1.9mm]}-{Stealth[length=2.6mm,width=1.9mm]},
        shorten <=1pt,
        shorten >=1pt
    },
    dataflow/.style={
        draw=muted,
        line width=0.95pt,
        line cap=round,
        line join=round,
        rounded corners=2.2pt,
        -{Stealth[length=2.5mm,width=1.8mm]},
        shorten <=1pt,
        shorten >=1pt
    },
    warnflow/.style={
        draw=risk,
        line width=0.95pt,
        dashed,
        line cap=round,
        line join=round,
        rounded corners=2.2pt,
        -{Stealth[length=2.5mm,width=1.8mm]},
        shorten <=1pt,
        shorten >=1pt
    }
]

\node[layer] (chanlbl) at (0,0) {External messaging channels};

\node[ext, below=4mm of chanlbl, xshift=-30mm] (wa) {WhatsApp};
\node[ext, right=3mm of wa] (tg) {Telegram};
\node[ext, right=3mm of tg] (dc) {Discord};
\node[ext, right=3mm of dc] (sl) {Slack};

\node[ifacebox, below=16mm of tg] (msg) {Messaging\\Interface};
\node[riskbox, below=7mm of msg] (gw) {Gateway\\[-0.4mm]\scriptsize(HTTP / WebSocket)};

\node[controlbox, below=8mm of gw] (ctrl) {LLM Controller\\[-0.4mm]\scriptsize(Cognitive Core)};
\node[controlbox, right=13mm of ctrl] (prov) {Provider\\Abstraction};

\node[layer, above=4mm of prov] (llmlbl) {Model backends};
\node[ext, above=3mm of llmlbl, minimum width=34mm] (llm)
    {LLM Providers\\[-0.4mm]\scriptsize(Claude / GPT / Gemini)};

\node[controlbox, below=8mm of ctrl] (appr) {Approval\\System};

\node[execbox, below left=10mm and 12mm of appr] (tools) {Tool Execution\\Engine};
\node[riskbox, below right=10mm and 12mm of appr] (skills) {Skill System\\[-0.4mm]\scriptsize(ClawHub)};

\node[riskbox, below=15mm of tools] (mem) {Persistent State\\[-0.4mm]\scriptsize(SOUL.md, MEMORY.md)};
\node[statebox, right=13mm of mem] (cred) {Credential Store\\[-0.4mm]\scriptsize(\texttt{\~/.openclaw/})};

\node[
    ext,
    below=15mm of $(mem)!0.5!(cred)$,
    minimum width=70mm,
    minimum height=9mm,
    font=\footnotesize
] (os)
    {Host Operating System\\[-0.4mm]\scriptsize(File System, Shell, Network)};

\begin{scope}[on background layer]
    \node[
        draw=line,
        rounded corners=6pt,
        fill=panel,
        inner xsep=9pt,
        inner ysep=12pt,
        fit=(msg)(gw)(ctrl)(prov)(appr)(tools)(skills)(mem)(cred)
    ] (daemon) {};
\end{scope}

\node[
    anchor=south west,
    align=left,
    font=\scriptsize\bfseries,
    text=muted,
    fill=white,
    inner xsep=2pt,
    inner ysep=1pt
] at ([xshift=2pt,yshift=1.5pt]daemon.north west)
{OpenClaw daemon\\monolithic single-process design};

\node[
    anchor=south east,
    font=\scriptsize\bfseries,
    text=muted,
    fill=white,
    inner xsep=2pt,
    inner ysep=1pt
] at ([xshift=-2pt,yshift=1.5pt]daemon.north east)
{ASTELD A1};

\node[rtag, anchor=south west] at ([xshift=1.0mm,yshift=0.8mm]gw.north east) {V1};
\node[rtag, anchor=south west] at ([xshift=1.0mm,yshift=0.8mm]appr.north east) {V2};
\node[rtag, anchor=south west] at ([xshift=1.0mm,yshift=0.8mm]tools.north east) {V3};
\node[rtag, anchor=south west] at ([xshift=1.0mm,yshift=0.8mm]skills.north east) {V4};
\node[rtag, anchor=south west] at ([xshift=1.0mm,yshift=0.8mm]cred.north east) {V5};

\coordinate (waIn) at ([xshift=-8.5mm]msg.north);
\coordinate (tgIn) at ([xshift=-2.8mm]msg.north);
\coordinate (dcIn) at ([xshift= 2.8mm]msg.north);
\coordinate (slIn) at ([xshift= 8.5mm]msg.north);

\coordinate (waTap) at ([yshift=6.5mm]waIn);
\coordinate (tgTap) at ([yshift=6.5mm]tgIn);
\coordinate (dcTap) at ([yshift=6.5mm]dcIn);
\coordinate (slTap) at ([yshift=6.5mm]slIn);

\draw[flow] (wa.south) -- ++(0,-3.6mm) -| (waTap) -- (waIn);
\draw[flow] (tg.south) -- ++(0,-3.6mm) -| (tgTap) -- (tgIn);
\draw[flow] (dc.south) -- ++(0,-3.6mm) -| (dcTap) -- (dcIn);
\draw[flow] (sl.south) -- ++(0,-3.6mm) -| (slTap) -- (slIn);

\draw[bflow] (msg.south) -- (gw.north);
\draw[bflow] (gw.south) -- (ctrl.north);
\draw[bflow] (ctrl.east) -- (prov.west);
\draw[flow] (prov.north) -- (llm.south);
\draw[flow] (ctrl.south) -- (appr.north);

\coordinate (toolsTap)  at ([yshift=4mm]tools.north);
\coordinate (skillsTap) at ([yshift=4mm]skills.north);

\draw[flow] ([xshift=-1.4mm]ctrl.south) -- ++(0,-3mm) -| (toolsTap) -- (tools.north);
\draw[flow] ([xshift= 1.4mm]ctrl.south) -- ++(0,-3mm) -| (skillsTap) -- (skills.north);

\coordinate (warnTap) at ([yshift=4mm]tools.north);
\draw[warnflow] ([xshift=-0.8mm]appr.west) -- ++(-6mm,0) |- (warnTap) -- (tools.north);

\draw[dataflow] (tools.south) -- (mem.north);

\coordinate (stateY)      at ([yshift=-4mm]cred.south);
\coordinate (skillsDrop)  at (skills.south |- stateY);
\coordinate (memLowIn)    at (mem.east |- stateY);
\coordinate (memRightLow) at ([xshift=6mm]memLowIn);
\coordinate (memRightIn)  at ([xshift=6mm]mem.east);

\draw[dataflow] (skills.south) -- (skillsDrop) -- (memRightLow) -- (memRightIn) -- (mem.east);

\draw[dataflow] (mem.east) -- (cred.west);

\coordinate (memOsIn)  at (mem.south |- os.north);
\coordinate (credOsIn) at (cred.south |- os.north);

\draw[dataflow] (mem.south)  -- (memOsIn);
\draw[dataflow] (cred.south) -- (credOsIn);

\node[layer, anchor=east] at ([xshift=-7mm]msg.west) {Ingress};
\node[layer, anchor=east] at ([xshift=-7mm]ctrl.west) {Orchestration};
\node[layer, anchor=east] at ([xshift=-7mm]tools.west) {Execution};
\node[layer, anchor=east] at ([xshift=-7mm]mem.west) {State};

\end{tikzpicture}

\caption{
OpenClaw architecture. Red-tinted nodes mark components with documented security issues. V1 denotes gateway exposure via \texttt{0.0.0.0}; V2 approval-bypass CVEs; V3 privileged tool execution; V4 the ClawHub skill attack surface; and V5 plaintext or unencrypted credential storage~\cite{microsoft2026running_openclaw}\cite{certik2026openclaw_report}\cite{zaitsev2026security_teams}.
}
\label{fig:architecture}
\end{figure}

\subsection{Architectural Overview}
OpenClaw follows a \textbf{monolithic daemon architecture} (ASTELD category A1):
a single long-running process that integrates the LLM controller, tool execution
engine, skill loader, messaging interface, and local gateway into one TypeScript
application. This design stands in contrast to the modular, microservice, or
graph-based architectures adopted by competing frameworks (see
Section~\ref{sec:ecosystem}).

The system comprises five principal components---Gateway, LLM Controller, Tool Execution Engine, Skill System, and Messaging Interface. Additional subcomponents (Provider Abstraction, Approval System, Persistent State, and Credential Store) are shown in Figure~\ref{fig:architecture} and discussed in Section~\ref{sec:security}.

\begin{enumerate}[leftmargin=*]
  \item \textbf{Gateway}: An HTTP/WebSocket server that exposes the agent's
  capabilities to local and remote clients. By default, the gateway binds to
  \texttt{localhost} on a dynamically assigned port, but misconfigurations
  frequently result in binding to \texttt{0.0.0.0}, exposing the agent to the
  public internet~\cite{microsoft2026running_openclaw}\cite{certik2026openclaw_report}\cite{zaitsev2026security_teams}.

  \item \textbf{LLM Controller}: The cognitive core that receives user
  instructions (via messaging platforms or the gateway), maintains conversational
  context, formulates plans, and selects tools for execution. OpenClaw supports
  multiple LLM backends---including Anthropic Claude, OpenAI GPT-4, Google
  Gemini, and local models via Ollama---through a provider abstraction layer
  shown as a distinct subcomponent in Figure~\ref{fig:architecture}~\cite{openclaw2026agent_runtime}.

  \item \textbf{Tool Execution Engine}: Responsible for executing the actions
  selected by the LLM controller. Tools include shell command execution, file
  system operations, HTTP requests, and skill invocations. The engine operates
  with the full privileges of the user account under which the daemon
  runs---typically the user's primary login account~\cite{microsoft2026running_openclaw}.

  \item \textbf{Skill System}: An extension mechanism that allows third-party
  developers to package tool definitions, prompt templates, and execution logic
  as installable ``skills.'' Skills are distributed through ClawHub, an open
  marketplace that had grown to 40,000+ entries by April~2026~\cite{gradually2026statistics}\cite{openclawvps2026statistics}. The
  skill system is the primary vector for the supply chain attacks described in
  Section~\ref{sec:security}.

  \item \textbf{Messaging Interface}: Adapters that connect the daemon to
  messaging platforms (WhatsApp, Telegram, Discord, Slack). This component is
  architecturally significant because it means user instructions arrive through
  channels designed for human-to-human communication, not for machine
  control---a design choice with profound security implications~\cite{microsoft2026running_openclaw}.
\end{enumerate}

\subsection{Execution Model}

OpenClaw employs a \textbf{single-agent sequential execution loop} (ASTELD category E1): observe $\to$ think $\to$ act $\to$ observe. The LLM controller receives an instruction, reasons about the appropriate action, invokes a tool, observes the result, and iterates until the goal is achieved or the user intervenes.

This execution model has three notable properties:

\textbf{Unbounded iteration.} Unlike frameworks that impose step limits or token budgets, OpenClaw's default configuration allows the agent to iterate indefinitely. Combined with system-level tool access, this means a single malicious instruction can trigger an arbitrarily long chain of privileged actions~\cite{suwansathit2026openclaw_vulnerabilities}.

\textbf{Approval system.} OpenClaw implements a command approval mechanism in which certain actions (e.g., shell commands, file deletions) are presented to the user for confirmation before execution. However, as documented in Section~\ref{sec:security}, this system suffers from TOCTOU (time-of-check-time-of-use) vulnerabilities: the command displayed to the user may differ from the command actually executed~\cite{shan2026claw_grip}\cite{jgamblin2026openclawcves}.

\textbf{``Allow always'' mode.} Users can configure the approval system to automatically approve all actions of a given type. Microsoft's security assessment found that the majority of users default to this configuration for convenience, effectively operating at autonomy level L4 (delegated autonomy with no per-action oversight) despite the system's design for L2 (guided autonomy with approval)~\cite{microsoft2026running_openclaw}. This behavioral pattern---users choosing convenience over security---is a recurring theme in the security literature on autonomous agents.

\subsection{The Lethal Trifecta}

As introduced in Section~\ref{sec:bg-security}, Microsoft's security analysis~\cite{microsoft2026running_openclaw}\cite{arnav2026prompt_injection} identifies a structural property of OpenClaw's architecture that it terms the \textbf{``lethal trifecta''}: the simultaneous presence of (1)~access to private data on the user's device, (2)~exposure to untrusted content from external sources (web pages, emails, documents), and (3)~the ability to communicate with and act upon external services.

When all three conditions hold, prompt injection transitions from a content moderation problem to an \textbf{authorization bypass}. An attacker who can embed a malicious instruction in a web page, email, or document that the agent processes can cause the agent to:

\begin{itemize}[leftmargin=*]
  \item Read and exfiltrate private files (condition~1 enables access, condition~3 enables exfiltration)
  \item Install software or modify system configurations (condition~3 enables action)
  \item Persist the attack by modifying the agent's memory files (SOUL.md, MEMORY.md), ensuring the malicious behavior survives restarts~\cite{shan2026claw_grip}\cite{penligent2026redteam}
\end{itemize}

This is not a bug in the traditional sense; it is an architectural property. The monolithic daemon design places the LLM controller, which processes both trusted instructions and untrusted content, in the same security domain as the tool execution engine, which has full system privileges. There is no privilege separation, no sandbox boundary, and no formal distinction between instruction and data at the architectural level~\cite{microsoft2026running_openclaw}\cite{certik2026openclaw_report}.

\subsection{State and Memory Management}

OpenClaw maintains several forms of persistent state:

\textbf{Conversational memory.} The LLM controller maintains a context window of recent interactions. This is ephemeral and bounded by the LLM's context length.

\textbf{Long-term memory.} OpenClaw supports persistent memory through two mechanisms: (1)~SOUL.md files that define the agent's personality and behavioral guidelines, and (2)~MEMORY.md files that store accumulated knowledge. Both are stored as plaintext Markdown files in the user's home directory (\texttt{\textasciitilde/.openclaw/}) and are loaded into the LLM's context at startup~\cite{shan2026claw_grip}.

\textbf{Credential storage.} API keys, tokens, and authentication credentials are stored in plaintext in \texttt{\textasciitilde/.openclaw/openclaw.json} and environment files. No encryption, no OS keychain integration, and no access controls beyond filesystem permissions~\cite{zaitsev2026security_teams}\cite{armo2026cve32922}.

The security implications of this state management approach are significant. Because SOUL.md and MEMORY.md are loaded into the LLM context, an attacker who can modify these files---through a prompt injection, a malicious skill, or direct file access---can persistently alter the agent's behavior. This ``memory poisoning'' attack vector is documented in Section~\ref{sec:security} and represents one of the most insidious threats to autonomous agent systems~\cite{shan2026claw_grip}\cite{penligent2026redteam}.

\subsection{Deployment Topology}

OpenClaw is designed as a \textbf{local-first personal agent} (ASTELD category D1): the daemon runs on the user's device, and the primary interaction occurs through messaging platforms. However, the reality of deployment diverges significantly from the design intent:

\begin{itemize}[leftmargin=*]
  \item \textbf{Approximately 500,000+ instances} were reported as internet-facing in late-March~2026 live checks, with \textbf{15,000+} directly exploitable via known remote-code-execution paths~\cite{securityscorecard2026attack_surface}\cite{columbus2026instances}.
  \item \textbf{More than 30,000 exposed instances} were observed with material security risks during the same reporting window~\cite{columbus2026instances}.
  \item The gateway, intended for localhost communication, is frequently exposed to the public internet due to Docker misconfiguration, cloud deployment, or explicit user choice~\cite{microsoft2026running_openclaw}.
\end{itemize}

This gap between designed topology (local-first, personal) and actual topology (internet-exposed, multi-user) is a recurring source of vulnerabilities. The security model assumes a trusted local environment; the deployment reality is an untrusted networked environment.

\subsection{Architectural Comparison via ASTELD}

To contextualize OpenClaw's architectural decisions, we apply the ASTELD classification framework introduced in this paper (see Section~\ref{sec:ecosystem} for full framework description and cross-platform comparison). OpenClaw's ASTELD coordinates are:

\begin{table}[tbp]
\centering
\caption{ASTELD profile of OpenClaw.}
\label{tab:openclaw_asteld_profile}
\footnotesize
\setlength{\tabcolsep}{4pt}
\renewcommand{\arraystretch}{1.06}
\begin{tabular}{@{}>{\RaggedRight\arraybackslash}p{2.95cm} >{\centering\arraybackslash}p{2.05cm} >{\RaggedRight\arraybackslash}p{5.85cm}@{}}
\toprule
\textbf{Axis} & \multicolumn{1}{c}{\textbf{Category}} & \textbf{Description} \\
\midrule
Architecture & A1 & Monolithic daemon \\
Security & S2 & Reactive, post-hoc patching \\
Tool Integration & T2 & Plugin marketplace (ClawHub) \\
Execution & E1 & Single-agent sequential loop \\
Level of autonomy & L2/L4 & Designed for L2; users default to L4 \\
Deployment & D1 & Local-first personal agent \\
\bottomrule
\end{tabular}
\end{table}

This profile is unique among the eight frameworks analyzed in Section~\ref{sec:ecosystem}. No other framework combines A1 (monolithic) with T2 (marketplace) and D1 (local-first). The combination explains both OpenClaw's distinctive strengths---simplicity, accessibility, low setup friction---and its distinctive vulnerabilities---no privilege separation, marketplace as attack surface, local daemon with internet exposure.

\subsection{Summary}
OpenClaw's architecture is a study in trade-offs. The monolithic daemon design
enables the simplicity and accessibility that drove viral adoption, but it also
creates a security surface that is fundamentally different from---and more
dangerous than---that of modular or graph-based frameworks. The lethal trifecta
is not a flaw to be patched; it is a consequence of architectural decisions that
prioritized user experience over security isolation. As we demonstrate in the
next section, the vulnerabilities that follow are largely predictable from this
architectural profile---rooted in design decisions rather than implementation
accidents.

\section{Security and Privacy Analysis}
\label{sec:security}

OpenClaw's security trajectory is unlike that of any prior open-source project. Within five months of its viral breakout, public trackers documented 100+ tracked advisories and 10+ published CVEs, while broader vendor summaries reported larger tallies of tracked vulnerability records under different aggregation rules; these sources also converged on the presence of several critical-severity vulnerabilities (CVSS $\geq$ 9.0), dozens of additional high-severity findings, and a 4-day burst in which 9 new security-tracker entries were logged between March~18 and March~21, 2026~\cite{suwansathit2026openclaw_vulnerabilities}\cite{jgamblin2026openclawcves}\cite{blink2026security}. At the time of writing, approximately 500,000+ OpenClaw instances had been reported as internet-facing, more than 30,000 were observed with material security risks, and 15,000+ were directly exploitable via known RCE paths~\cite{securityscorecard2026attack_surface}\cite{columbus2026instances}.

This section presents a six-category vulnerability taxonomy synthesized from CVE databases, five institutional security assessments~\cite{microsoft2026running_openclaw}\cite{certik2026openclaw_report}\cite{zaitsev2026security_teams}\cite{tucci2026cisos}\cite{koi2026clawhavoc}, and academic research~\cite{suwansathit2026openclaw_vulnerabilities}\cite{shan2026claw_grip}\cite{wang2026double_agent}. We then identify four canonical attack chains that compose vulnerabilities across categories, and conclude with an assessment of the defense landscape.

\begin{table}[tbp]
\centering
\caption{OpenClaw security vulnerability taxonomy.}
\label{tab:security}

\footnotesize
\setlength{\tabcolsep}{4pt}
\renewcommand{\arraystretch}{1.08}

\newcolumntype{C}{>{\centering\arraybackslash}p{0.9cm}}
\newcolumntype{N}{>{\RaggedRight\arraybackslash}p{2.25cm}}
\newcolumntype{R}{>{\RaggedRight\arraybackslash}p{3.15cm}}
\newcolumntype{E}{>{\RaggedRight\arraybackslash}X}
\newcolumntype{S}{>{\centering\arraybackslash}p{1.45cm}}
\newcolumntype{T}{>{\centering\arraybackslash}p{1.18cm}}

\begin{tabularx}{\linewidth}{@{}C N R E S T@{}}
\toprule
\textbf{Cat.} & \textbf{Category Name} & \textbf{Root Cause} & \textbf{Representative Evidence} & \textbf{CVSS / Severity} & \textbf{Count} \\
\midrule

A & Authentication \& Session Trust Abuse &
Trust-boundary failures; improper scope/token binding in WebSocket-based control paths &
CVE-2026-25253 (ClawJacked)\newline
CVE-2026-22172 (Scope Self-Decl.)\newline
CVE-2026-32922 (Pairing Escalation) &
8.8--9.9\newline
High--Crit. &
3+ \\

\addlinespace[2pt]
B & Approval System Bypass &
Display-time vs.\ execution-time inconsistency (TOCTOU analog) &
CVE-2026-29607 (Allow-Always)\newline
CVE-2026-28460 (Line Continuation)\newline
CVE-2026-34426 (Env Var Injection)\newline
CVE-2026-32065 (Token Mismatch) &
6.5--8.8\newline
Med.--High &
4 \\

\addlinespace[2pt]
C & OS-Specific Escapes &
Shell escaping failures across platforms &
CVE-2026-22179 (macOS cmd subst.)\newline
CVE-2026-22176 (Windows .cmd) &
High\newline
7.8 &
2 \\

\addlinespace[2pt]
D & Prompt Injection &
No architectural separation of instructions from data in LLM context &
Direct, indirect (IDPI/XPIA), memory poisoning, guidance injection, cross-agent propagation &
--- &
5 sub-types \\

\addlinespace[2pt]
E & Supply Chain (ClawHavoc) &
Unvetted marketplace; weak publisher vetting; unvetted publishing pipeline &
1{,}184 malicious skills identified\newline
Trojan/OpenClaw.\allowbreak PolySkill &
--- &
1{,}184 skills \\

\addlinespace[2pt]
F & Infrastructure Misconfig. &
Local-dev defaults deployed at internet scale &
CVE-2026-32018 (No file locking)\newline
500K+ internet-facing instances\newline
15K+ known-RCE exploitable\newline
Plaintext credential storage &
6.6\newline
---\newline
--- &
3+ \\

\bottomrule
\end{tabularx}
\vspace{2pt}

\raggedright\footnotesize
Representative evidence items include CVEs where assigned and non-CVE records where the category is documented through tracker entries, marketplace audits, or infrastructure exposure reports.
\end{table}

\subsection{Vulnerability Taxonomy}

We classify OpenClaw vulnerabilities into six categories based on attack vector, impact, and architectural root cause. The categories are not mutually exclusive: real-world attacks frequently chain vulnerabilities across categories (see Section~\ref{sec:attack-chains}).

\subsubsection{Category A: Authentication and Session Trust Abuse}

The architectural root cause is OpenClaw's trust assumption that all localhost connections are legitimate---an assumption that does not hold when the gateway is internet-exposed or when browser-based attacks target localhost.

Three critical CVEs define this category. CVE-2026-25253 (ClawJacked/ClawBleed, CVSS~8.8) enables a malicious webpage to initiate cross-site WebSocket hijacking against the localhost gateway; because the browser's Same-Origin Policy does not block WebSocket connections to localhost, and OpenClaw's rate limiter exempts 127.0.0.1, an attacker can brute-force the gateway port and exfiltrate authentication tokens, achieving full agent takeover~\cite{oasis2026clawjacked}\cite{hackernews2026clawjacked}. CVE-2026-22172 (CVSS~9.9) allows a WebSocket client to self-declare administrative scopes, bypassing authentication entirely---the gateway trusts the client's scope assertion without verification~\cite{jgamblin2026openclawcves}. CVE-2026-32922 (Critical) enables a single API call to escalate a pairing token into full administrative control with remote code execution capabilities~\cite{armo2026cve32922}.

The common thread is a \textbf{trust boundary violation}: mechanisms designed for trusted local environments are deployed in untrusted networked environments.

\subsubsection{Category B: Approval System Bypass}

OpenClaw's command approval system is its primary human-in-the-loop security mechanism---and it has been bypassed through four distinct CVEs, each exploiting a different inconsistency between the approval display path and the execution path.

CVE-2026-29607 exploits the fact that ``allow always'' persists at the wrapper command level, not the inner command; an attacker can swap the inner payload after the wrapper has been approved, achieving persistent RCE without re-prompting~\cite{jgamblin2026openclawcves}. CVE-2026-28460 demonstrates that shell line-continuation characters bypass the command allowlist entirely~\cite{jgamblin2026openclawcves}. CVE-2026-34426 shows that inconsistent environment variable normalization between the approval and execution paths enables environment variable injection without triggering the approval prompt~\cite{jgamblin2026openclawcves}. CVE-2026-32065 (CVSS~5.7) reveals a mismatch between how command tokens are displayed during approval and how they are resolved at execution time~\cite{shan2026claw_grip}.

The architectural insight is that the approval system suffers from a class of vulnerability analogous to the TOCTOU (time-of-check-time-of-use) problem in operating system security~\cite{shan2026claw_grip}. The ``check'' (displaying the command for approval) and the ``use'' (executing the command) operate on different representations of the same action, creating a semantic gap that attackers can exploit. No formal specification of approval correctness exists for any agent system, making this an open research problem (see RQ3, Section~\ref{sec:rq}).

\subsubsection{Category C: Operating System-Specific Escapes}

Platform-specific shell escaping failures create additional attack vectors. CVE-2026-22179 (High) allows command substitution syntax inside double-quoted strings to bypass the command allowlist on macOS~\cite{jgamblin2026openclawcves}. CVE-2026-22176 (CVSS~7.8) exploits unescaped environment variables in scheduled task \texttt{.cmd} scripts on Windows, where characters such as \texttt{\&}, \texttt{|}, and \texttt{\^{}} are interpreted as command separators~\cite{jgamblin2026openclawcves}.

These vulnerabilities, while platform-specific, illustrate a broader point: OpenClaw's tool execution engine must correctly handle shell semantics across every supported operating system---a combinatorially complex problem that is intrinsically difficult to solve in a monolithic architecture where the LLM controller generates shell commands directly.

\subsubsection{Category D: Prompt Injection}

Prompt injection in autonomous agents is qualitatively different from prompt injection in chatbots. In a chatbot, a successful injection may cause the model to generate inappropriate text. In an agent with system-level tool access, a successful injection can cause the execution of arbitrary shell commands, the exfiltration of private data, and the persistent modification of the agent's behavior.

We identify five sub-types of prompt injection relevant to OpenClaw:

\textbf{Direct injection}: The user types malicious instructions that override the system prompt, causing the agent to execute unintended operations~\cite{suwansathit2026openclaw_vulnerabilities}.

\textbf{Indirect injection (IDPI/XPIA)}: Malicious instructions are embedded in content---web pages, emails, documents, MCP tool outputs---that the agent processes during legitimate task execution. Because the LLM processes instructions and data in the same context window with no architectural separation, the embedded instructions can hijack the agent's reasoning~\cite{shan2026claw_grip}\cite{penligent2026redteam}\cite{esecurityplanet2026prompt_injection}.

\textbf{Memory poisoning}: An attacker modifies the agent's persistent memory files (SOUL.md, MEMORY.md) to inject instructions that alter the agent's behavior. Because these files are loaded into the LLM context at every startup, the injected behavior persists across restarts and may include time-delayed activation triggers~\cite{shan2026claw_grip}\cite{penligent2026redteam}. This attack vector is unique to agent systems with persistent state and represents one of the most insidious threats identified in the literature.

\textbf{Guidance injection}: Attacker-controlled instructions in ClawHub skill descriptions or MCP server tool outputs alter the agent's reasoning during skill execution. Unlike indirect injection through user data, guidance injection operates through the tool layer itself---the agent trusts tool outputs as authoritative, creating a privilege escalation path~\cite{repello2026clawhavoc}\cite{koi2026clawhavoc}.

\textbf{Cross-agent propagation}: In multi-instance deployments or social-agent networks (such as Moltbook), a malicious prompt in shared content can reach multiple agents simultaneously, causing coordinated compromise~\cite{openclaw2026agent_runtime}\cite{manik2026moltbook}. This vector remains under-studied; only one empirical analysis exists~\cite{openclaw2026agent_runtime}, and no formal threat model for agent-to-agent infection has been proposed (see RQ6, Section~\ref{sec:rq}).

The key insight, shared by Microsoft~\cite{microsoft2026running_openclaw}, CrowdStrike~\cite{zaitsev2026security_teams}, and academic researchers~\cite{suwansathit2026openclaw_vulnerabilities}\cite{shan2026claw_grip}, is that prompt injection in agentic systems is an \textbf{authorization problem}, not a content moderation issue. Tool access is the amplifier: systems without external tool access show minimal successful injection outcomes. The architectural separation of instruction processing from data processing remains unsolved (see RQ1, Section~\ref{sec:rq}).

\subsubsection{Category E: Supply Chain Attacks (ClawHavoc)}

The ClawHavoc campaign, disclosed in February~2026, represents the first large-scale supply chain attack targeting an AI agent skill marketplace~\cite{cyberpress2026clawhavoc}\cite{repello2026clawhavoc}\cite{koi2026clawhavoc}. The campaign's root cause was ClawHub's minimal vetting requirement: publishing a skill required only a GitHub account that was one week old.

The attack was executed through multiple sub-campaigns. The primary ClawHavoc campaign used fake error messages to trick users into pasting base64-encoded commands that installed Atomic Stealer (AMOS)---malware that harvests browser credentials, cryptocurrency wallets, and system data~\cite{cyberpress2026clawhavoc}. The AuthTool sub-campaign deployed dormant payloads activated by specific user prompts, establishing persistent reverse shells~\cite{repello2026clawhavoc}. The Hidden Backdoor campaign disguised itself as an Apple Software Update during skill installation, creating an encrypted tunnel to attacker infrastructure~\cite{koi2026clawhavoc}. A benign-looking weather assistant skill was found to steal API keys from the \texttt{.clawdbot/.env} file~\cite{koi2026clawhavoc}.

The scale of the campaign is striking. Koi Security's initial audit identified 341 malicious skills among 2,857 total (approximately 12\%)~\cite{cyberpress2026clawhavoc}. Bitdefender's expanded scan found 824+ among 10,700+ skills~\cite{repello2026clawhavoc}. Subsequent reporting later placed the total at 1,184 malicious skills, and Koi Security assigned the classification Trojan/OpenClaw.PolySkill~\cite{cyberpress2026clawhavoc}\cite{koi2026clawhavoc}. A single threat actor (``hightower6eu'') was responsible for 354 of these skills~\cite{cyberpress2026clawhavoc}.

ClawHavoc established a new threat class: \textbf{AI agent supply chain attacks through natural language payloads}. Unlike traditional software supply chain attacks (e.g., npm, PyPI package poisoning) where malicious code can be detected through static analysis, agent skill poisoning can operate entirely through natural language instructions that the LLM interprets as commands. This makes conventional code scanning tools, including VirusTotal (which OpenClaw adopted post-ClawHavoc), fundamentally insufficient for detection~\cite{koi2026clawhavoc}\cite{arnav2026prompt_injection}.

\subsubsection{Category F: Infrastructure Misconfigurations}

Default configurations designed for local development are routinely deployed at internet scale. The gateway binds to \texttt{0.0.0.0} (exposing all network interfaces) in many deployment configurations; by late March~2026, public reporting described approximately 500,000+ internet-facing instances, more than 30,000 with material security risks, and 15,000+ directly exploitable through known RCE paths~\cite{securityscorecard2026attack_surface}\cite{columbus2026instances}. API keys and credentials are stored in plaintext in \texttt{\textasciitilde/.openclaw/openclaw.json} and \texttt{.env} files with no encryption or OS keychain integration~\cite{zaitsev2026security_teams}\cite{armo2026cve32922}. Authentication tokens transmitted in URL query parameters are leaked to browser history and server logs~\cite{jgamblin2026openclawcves}. The absence of file locking (CVE-2026-32018, CVSS~6.6) causes concurrent registry operations to produce orphaned containers and inconsistent state~\cite{jgamblin2026openclawcves}.

\subsection{Attack Chain Taxonomy}
\label{sec:attack-chains}

Individual vulnerabilities rarely exist in isolation. We identify four canonical attack chains that combine vulnerabilities across the categories defined above.

\textbf{Chain~1: Browser-Initiated Full Takeover.} A malicious webpage (Category~A: ClawJacked) initiates a WebSocket brute-force against the victim's localhost gateway $\to$ exfiltrates the authentication token $\to$ achieves full agent control $\to$ executes arbitrary shell commands. This chain requires only that the victim visits a webpage while OpenClaw is running---no social engineering, no malware installation, no user interaction beyond the initial page visit~\cite{oasis2026clawjacked}.

\textbf{Chain~2: Supply Chain Persistent Compromise.} A malicious ClawHub skill (Category E) installs and tampers with persistent instruction files such as SOUL.md or MEMORY.md (Category D: memory poisoning), creating a delayed behavioral backdoor. When the trigger condition is later met, the poisoned memory can steer the agent toward credential exfiltration or attacker-directed command execution; the compromise persists across restarts because the poisoned memory is reloaded at startup~\cite{repello2026clawhavoc}\cite{koi2026clawhavoc}.

\textbf{Chain~3: Indirect Injection Escalation.} A poisoned web page, email, or document (Category~D: indirect injection) is ingested by the agent during legitimate task execution $\to$ the embedded instructions override the agent's reasoning $\to$ the agent exploits an approval bypass (Category~B) or executes a shell command directly $\to$ data is exfiltrated to an attacker-controlled server~\cite{shan2026claw_grip}\cite{penligent2026redteam}.

\textbf{Chain~4: Cross-Agent Propagation.} A poisoned message introduced into a shared feed or Moltbook interaction (Category D: cross-agent exposure) can be processed by multiple agent instances. If those agents are configured to write untrusted content into persistent memory, the same malicious instruction may propagate into separate memory states, creating the potential for repeated unauthorized actions across multiple agents~\cite{openclaw2026agent_runtime}.

\subsection{Institutional Security Assessments}

The convergence of institutional assessments regarding OpenClaw is noteworthy. Five major security organizations independently evaluated the platform and reached remarkably consistent conclusions, suggesting that its vulnerabilities are systemic rather than superficial.

Microsoft Security emphasized the risks to host environments, recommending that users avoid deploying OpenClaw with primary work or personal accounts and proposing the use of dedicated virtual machines with non-privileged credentials as a minimum precaution~\cite{microsoft2026running_openclaw}. CertiK identified a fundamental architectural contradiction, characterizing the vulnerability profile as a flaw designed for trusted local use, but deployed at internet scale, and advocated for mobile-OS-style permission declarations for all integrated skills~\cite{certik2026openclaw_report}.

From a threat-vector perspective, CrowdStrike classified OpenClaw as an AI super agent requiring intensive security scrutiny, warning that its extensive tool access serves as a primary amplifier for prompt injection attacks~\cite{zaitsev2026security_teams}. Trend Micro focused on enterprise governance, advising Chief Information Security Officers (CISOs) to apply the same level of security rigor to OpenClaw deployments as is standard for production-grade servers~\cite{tucci2026cisos}. Complementing these views, Koi Security provided critical insights into supply chain integrity, documenting the ClawHavoc activity and the proliferation of malicious skills designed to exploit the platform's execution model~\cite{koi2026clawhavoc}.

The institutional consensus is definitive: OpenClaw's security challenges are structural, not incidental. These vulnerabilities are direct consequences of foundational architectural decisions---specifically the monolithic daemon design, the unvetted marketplace, plaintext credential storage, and the localhost trust assumption---rather than implementation bugs that can be resolved through isolated patches.
\subsection{Defense Landscape}

The defense landscape spans three tiers of intervention:

\textbf{Tier~1: OpenClaw's own mitigations.} Post-disclosure releases addressed several concrete flaws, including the one-click RCE path fixed in version~2026.1.29, later hardening of gateway authentication and localhost handling, and subsequent closures for multiple March-reported attack paths~\cite{openclawpulse2026architecture}\cite{jgamblin2026openclawcves}. After ClawHavoc, the ecosystem also moved toward malware screening and user reporting for skill submissions~\cite{jgamblin2026openclawcves}. These mitigations are primarily reactive: they reduce exposure to known exploit paths, but do not by themselves eliminate the underlying architectural weaknesses.

\textbf{Tier~2: Community and derivative approaches.} NanoClaw adopts per-agent containerized isolation, limiting each agent's access to explicitly mounted resources~\cite{ossinsight2026forks_wave}. Other derivative or adjacent systems propose stronger guardrails such as owner-governed policy controls, sandboxed execution, and architectural separation between model interpretation and tool invocation~\cite{esecurityplanet2026prompt_injection}\cite{ossinsight2026forks_wave}.

\textbf{Tier~3: Third-party security tools.} Third-party defenses aim to monitor and constrain agent behavior at runtime. For example, Prompt Security's AI Gateway inspects traffic between AI applications and MCP servers and enforces allow/block policies on risky interactions~\cite{esecurityplanet2026prompt_injection}. Enterprise hardening guidance further recommends container or VM isolation, secrets minimization, memory-file integrity monitoring, and outbound connection monitoring for exfiltration detection~\cite{microsoft2026running_openclaw}\cite{zaitsev2026security_teams}.

However, these defenses mostly reduce exploitability or blast radius rather than removing the deeper failure modes identified in our taxonomy, including localhost trust-boundary violations, approval-time vs. execution-time semantic mismatch, and the lack of reliable separation between instructions and untrusted data in agent context.

\subsection{Summary}

OpenClaw's security landscape reveals a fundamental mismatch between deployment scale and security maturity. The six vulnerability categories---authentication hijacking, approval bypass, OS-specific escapes, prompt injection, supply chain attacks, and infrastructure misconfigurations---are not independent; they compose into multi-stage attack chains that amplify individual vulnerabilities into full system compromise. The institutional consensus, the volume of CVEs, and the architectural analysis all point to the same conclusion: the challenges are systemic, rooted in architectural decisions that cannot be fully remediated through patching alone. As we show in the next section, these challenges are not unique to OpenClaw but are shared, to varying degrees, across the autonomous agent ecosystem.

\section{ASTELD: Classification Framework and Empirical Evaluation}
\label{sec:ecosystem}

This section presents ASTELD as the paper's primary methodological contribution, specifies its construction and assignment protocol, and evaluates it on eight autonomous AI agent platforms. OpenClaw is then used as the anchor case for connecting classification coordinates to security, adoption, and derivative-ecosystem outcomes.

\subsection{Problem Setting and Scope}

The autonomous AI agent ecosystem has expanded rapidly since AutoGPT's viral release in March~2023~\cite{github_autogpt}. By April~2026, eight major open-source frameworks compete across different architectural philosophies, security models, and target use cases: OpenClaw~\cite{openclaw2026agent_runtime}, Hermes~\cite{github_hermes}, AutoGPT~\cite{github_autogpt}, CrewAI~\cite{agenttimes2026crewai}, LangGraph~\cite{langchain2025langgraph}, n8n~\cite{n8n2026platform}, Dify~\cite{dify2025milestone}, and AutoGen/Microsoft Agent Framework~\cite{autogen2023framework}\cite{microsoftlearn2026agent_framework}. These frameworks differ significantly in architecture, tool integration, level-of-autonomy mechanisms, and deployment strategies, making direct comparison challenging. Hermes is especially relevant because it targets the same broad personal-agent space as OpenClaw but shifts the deployment model toward a server-resident D2 profile and emphasizes a built-in learning loop with persistent cross-session memory~\cite{github_hermes}. While prior work has explored aspects of these systems, there remains a need for a more standardized, multi-dimensional comparison. Prior analyses have provided valuable insights through framework-specific studies~\cite{suwansathit2026openclaw_vulnerabilities}\cite{shan2026claw_grip}\cite{preprints2026openclaw_survey}, pairwise comparisons~\cite{letsdatascience2026langgraph}\cite{openclawai2026comparison}, and practical rankings or benchmarks~\cite{nocobase2026top20}, but they typically focus on specific perspectives or evaluation criteria. To address this need, ASTELD treats platform comparison as a categorical coding problem: observable evidence is mapped to a six-coordinate profile under explicit assignment rules, allowing systems to be compared without reducing them to a single ranking.

\subsection{Framework Construction and Evaluation Protocol}

ASTELD was developed through a four-stage procedure. First, we synthesized dimensions from prior agent, autonomy, security, and leakage taxonomies and retained dimensions that describe platform-level design rather than task-level performance. Second, candidate axes were screened for \emph{observability} (values can be assigned from documentation or implementation evidence), \emph{discrimination} (the axis separates at least two systems), \emph{non-redundancy} (the axis captures information not subsumed by another axis), and \emph{extensibility} (new platforms can be added without changing the coordinate system). Third, we defined categorical anchors for each axis and applied a dominant-configuration rule: a platform receives the category that best represents its default or most typical use, while ranges or transition notation are retained when common deployments materially span categories. Fourth, we evaluated the framework at three levels: profile discrimination across eight platforms, structural utility through cross-axis pattern analysis, and explanatory utility through the direction of change observed in 50+ OpenClaw derivatives.

The unit of analysis is the platform rather than an individual agent run, application, or optional component. Evidence is drawn from platform documentation, repositories, reported feature sets, and the security and ecosystem sources analyzed in the surrounding case study. ASTELD is therefore an empirical classification framework, not a performance benchmark: it organizes design properties and supports comparative reasoning, but does not claim that a higher category number is universally better or that the observed relationships are causal.

\subsection{ASTELD Axes and Category Definitions}

We propose ASTELD, a six-axis taxonomy for classifying autonomous AI agent platforms, where each axis is defined by observable system properties and assigned using explicit criteria. The name is an acronym for its axes: \textbf{A}rchitecture pattern, \textbf{S}ecurity posture, \textbf{T}ool integration model, \textbf{E}xecution paradigm, \textbf{L}evel of autonomy and human control, and \textbf{D}eployment topology. The framework was designed to satisfy four properties: (1)~\textbf{discriminative}---every axis separates at least two frameworks, and frameworks can be consistently mapped to distinct dominant profiles even when some axes require range-valued entries under common configurations; (2)~\textbf{comprehensive}---the axes jointly capture key technical, security, and operational properties; (3)~\textbf{empirically grounded}---every axis value is derived from observed framework properties, not theoretical design space; and (4)~\textbf{extensible}---derivatives and future frameworks can be positioned without restructuring the taxonomy.

ASTELD builds upon and extends six prior taxonomies identified in the literature, each of which addresses a subset of the classification problem: Sapkota et al.'s conceptual distinction between AI agents and agentic AI~\cite{sapkota2025agentic_ai}, the six-component agent anatomy of V et al.~\cite{v2026agentic_ai}, Abou Ali et al.'s dual-paradigm framework~\cite{abouali2025agentic_ai}, Interface EU's five-level autonomy classification~\cite{interfaceeu2024autonomy}, Piccialli et al.'s industry-focused autonomy progression~\cite{piccialli2025agentai}, and the Frontiers data leakage taxonomy~\cite{frontiers2026data_leakage}.

The six axes are defined as categorical dimensions. Each framework is assigned a dominant value per axis where possible; when common real-world configurations materially span multiple categories, we retain an explicit range or transition notation rather than forcing an artificial single-value assignment.

\textbf{Axis~1: Architecture Pattern (A).} Classifies the structural organization of the agent system into five categories: A1 (Monolithic Daemon)---a single long-running process handling all agent functions; A2 (Visual Workflow)---node-graph editor as primary interface with execution following a visual DAG; A3 (Composable Library)---agent primitives provided as a library for developer composition; A4 (Graph-Based Runtime)---stateful directed graph with cycles, checkpointing, and replay; A5 (Event-Driven Multi-Agent)---event bus architecture with asynchronous message passing across multiple agents.

\textbf{Axis~2: Security Posture (S).} Classifies the maturity of built-in security mechanisms, adapted from software security maturity models (BSIMM, OWASP SAMM), into four levels: S1 (Minimal)---no built-in security boundary, relying on OS-level permissions; S2 (Reactive)---post-hoc patches, approval prompts, allowlists; S3 (Framework-Level)---middleware permissions, tool-level access control, code execution sandboxing; S4 (Enterprise-Grade)---RBAC, SSO/SAML, audit logging, SIEM integration, air-gapped deployment support.

\textbf{Axis~3: Tool Integration Model (T).} Classifies how agents discover and invoke external tools: T1 (Static Configuration)---tools predefined at setup; T2 (Plugin Marketplace)---install-time selection from a curated or open marketplace; T3 (Protocol-Based MCP)---standardized Model Context Protocol for runtime tool discovery and invocation; T4 (Connector Ecosystem)---large library of pre-built connectors in an iPaaS-style model.

\textbf{Axis~4: Execution Paradigm (E).} Classifies computation structure: E1 (Single-Agent Loop)---observe-think-act cycle with sequential execution; E2 (Multi-Agent Conversation)---structured message exchange among multiple agents; E3 (Stateful Graph Execution)---state machine with branching, loops, and checkpoints; E4 (Event-Driven Pipeline)---trigger-node chain with workflow-level parallelism.

\textbf{Axis~5: Level of autonomy and human control (L).} Classifies the level of autonomous decision authority and human involvement, adapted from Interface EU's five-level model~\cite{interfaceeu2024autonomy}: L1 (Tool Mode)---every action requires explicit user instruction; L2 (Guided Autonomy)---the agent proposes actions, and a human approves each one; L3 (Bounded Autonomy)---the agent executes within a predefined scope; L4 (Delegated Autonomy)---the agent decomposes and executes complex goals end-to-end; L5 (Full Autonomy)---continuous operation without human oversight. To avoid conflict with the first Architecture Pattern (A), the sublevels of level of autonomy and human control use the prefix L (Level).

\textbf{Axis~6: Deployment Topology (D).} Classifies where the agent runtime executes: D1 (Local-First Personal)---runs on user's device with local file/system access; D2 (Self-Hosted Server)---runs on user-managed server; D3 (Cloud-Managed)---vendor-hosted agent-as-a-service; D4 (Library-Embedded)---agent logic embedded in user's application code.

\begin{figure}[tbp]
\centering
\definecolor{ink}{HTML}{223042}
\definecolor{line}{HTML}{55667A}
\definecolor{muted}{HTML}{738295}
\definecolor{panel}{HTML}{F7F9FC}
\definecolor{iface}{HTML}{EAF4FF}
\definecolor{control}{HTML}{ECF8F1}
\definecolor{exec}{HTML}{F3EEFF}
\definecolor{state}{HTML}{FFF7EA}
\definecolor{riskfill}{HTML}{FDEEEF}
\definecolor{risk}{HTML}{C75454}

\resizebox{0.98\textwidth}{!}{
\begin{tikzpicture}[
    x=1cm, y=1cm,
    every node/.style={align=center},
    title/.style={
        rectangle, rounded corners=3pt,
        draw=line, fill=iface,
        font=\small\bfseries, text=ink,
        minimum width=3.2cm, minimum height=0.7cm,
        inner sep=3pt
    },
    axishead/.style={
        rectangle, rounded corners=2pt,
        draw=line, fill=state,
        font=\scriptsize\bfseries, text=ink,
        minimum width=2.4cm, minimum height=0.6cm,
        inner sep=2pt
    },
    catbox/.style={
        rectangle, rounded corners=2pt,
        draw=muted!60, fill=white,
        font=\scriptsize, text=ink,
        text width=2.3cm,
        minimum height=0.6cm,
        inner sep=2pt
    }
]

\node[font=\normalsize\bfseries, text=ink, anchor=north west] at (-0.5, 0.6) {(a)};

\node[title] at (7.5, 0) {ASTELD Framework};

\node[axishead] at (0, -1.3)  {A: Architecture};
\node[axishead] at (3, -1.3)  {S: Security};
\node[axishead] at (6, -1.3)  {T: Tool Integration};
\node[axishead] at (9, -1.3)  {E: Execution};
\node[axishead] at (12, -1.3) {L: Level of autonomy};
\node[axishead] at (15, -1.3) {D: Deployment};

\node[catbox] at (0, -2.4) {A1: Monolithic Daemon};
\node[catbox] at (0, -3.3) {A2: Visual Workflow};
\node[catbox] at (0, -4.2) {A3: Composable Library};
\node[catbox] at (0, -5.1) {A4: Graph Runtime};
\node[catbox] at (0, -6.0) {A5: Event-Driven};

\node[catbox] at (3, -2.4) {S1: Minimal};
\node[catbox] at (3, -3.3) {S2: Reactive};
\node[catbox] at (3, -4.2) {S3: Framework-Level};
\node[catbox] at (3, -5.1) {S4: Enterprise};

\node[catbox] at (6, -2.4) {T1: Static};
\node[catbox] at (6, -3.3) {T2: Marketplace};
\node[catbox] at (6, -4.2) {T3: MCP};
\node[catbox] at (6, -5.1) {T4: Connectors};

\node[catbox] at (9, -2.4) {E1: Single-Agent};
\node[catbox] at (9, -3.3) {E2: Multi-Agent};
\node[catbox] at (9, -4.2) {E3: Stateful Graph};
\node[catbox] at (9, -5.1) {E4: Event-Driven};

\node[catbox] at (12, -2.4) {L1: Tool};
\node[catbox] at (12, -3.3) {L2: Guided};
\node[catbox] at (12, -4.2) {L3: Bounded};
\node[catbox] at (12, -5.1) {L4: Delegated};
\node[catbox] at (12, -6.0) {L5: Full};

\node[catbox] at (15, -2.4) {D1: Local};
\node[catbox] at (15, -3.3) {D2: Self-Hosted};
\node[catbox] at (15, -4.2) {D3: Cloud};
\node[catbox] at (15, -5.1) {D4: Embedded};

\end{tikzpicture}
}
\vspace{8pt}

\resizebox{0.97\textwidth}{!}{
\begin{tikzpicture}[
    scale=0.98,
    every node/.style={font=\scriptsize},
]

\node[font=\small\bfseries, text=ink, anchor=north west] at (-4.5, 4.2) {(b)};

\def\numaxes{6}
\def\maxval{4}

\foreach \level in {1,2,3,4} {
    \draw[muted!25]
        (90:\level) -- (30:\level) -- (-30:\level) --
        (-90:\level) -- (-150:\level) -- (150:\level) -- cycle;
}

\foreach \a in {1,...,\numaxes} {
    \draw[muted!40] (0,0) -- ({90 - (\a-1)*360/\numaxes}:\maxval+0.5);
}

\node[anchor=south, font=\small\bfseries, text=ink] at (90:\maxval+0.7) {Architecture (A)};
\node[anchor=south west, font=\small\bfseries, text=ink] at (30:\maxval+0.7) {Security (S)};
\node[anchor=north west, font=\small\bfseries, text=ink] at (-30:\maxval+0.7) {Tool Integ.\ (T)};
\node[anchor=north, font=\small\bfseries, text=ink] at (-90:\maxval+1.0) {Execution (E)};
\node[anchor=north east, font=\small\bfseries, text=ink] at (-150:\maxval+0.7) {Level of autonomy (L)};
\node[anchor=south east, font=\small\bfseries, text=ink] at (150:\maxval+0.7) {Deployment (D)};

\foreach \level/\lbl in {1/1, 2/2, 3/3, 4/4} {
    \node[font=\tiny, anchor=east, text=muted] at ({90}:\level) {\lbl};
}

\draw[thick, line, fill=iface, fill opacity=0.45]
    (90:1) -- (30:2) -- (-30:2) -- (-90:1) -- (-150:4) -- (150:1) -- cycle;
\foreach \coord in {(90:1), (30:2), (-30:2), (-90:1), (-150:4), (150:1)} {
    \fill[line] \coord circle (2pt);
}

\draw[thick, control!70!black, fill=control, fill opacity=0.35]
    (90:2) -- (30:4) -- (-30:4) -- (-90:4) -- (-150:3) -- (150:2) -- cycle;
\foreach \coord in {(90:2), (30:4), (-30:4), (-90:4), (-150:3), (150:2)} {
    \fill[control!70!black] \coord circle (2pt);
}

\draw[thick, risk, fill=riskfill, fill opacity=0.35]
    (90:3) -- (30:3) -- (-30:3) -- (-90:2) -- (-150:3) -- (150:4) -- cycle;
\foreach \coord in {(90:3), (30:3), (-30:3), (-90:2), (-150:3), (150:4)} {
    \fill[risk] \coord circle (2pt);
}

\draw[thick, line] (-4.0, -4.8) -- (-3.3, -4.8);
\node[anchor=west] at (-3.2, -4.8) {OpenClaw};

\draw[thick, control!70!black] (-4.0, -5.3) -- (-3.3, -5.3);
\node[anchor=west] at (-3.2, -5.3) {n8n};

\draw[thick, risk] (-4.0, -5.8) -- (-3.3, -5.8);
\node[anchor=west] at (-3.2, -5.8) {CrewAI};

\draw[<->, thick, dashed, muted] (30:2.1) -- (30:3.9);
\node[font=\tiny, text=muted, anchor=west] at ($(30:3)+(.3,0)$) {Security gap};

\end{tikzpicture}
}

\caption{ASTELD taxonomy and representative profiles. Panel (a) lists the axes and categories. Panel (b) plots OpenClaw, n8n, and CrewAI.}
\label{fig:asteld}
\end{figure}
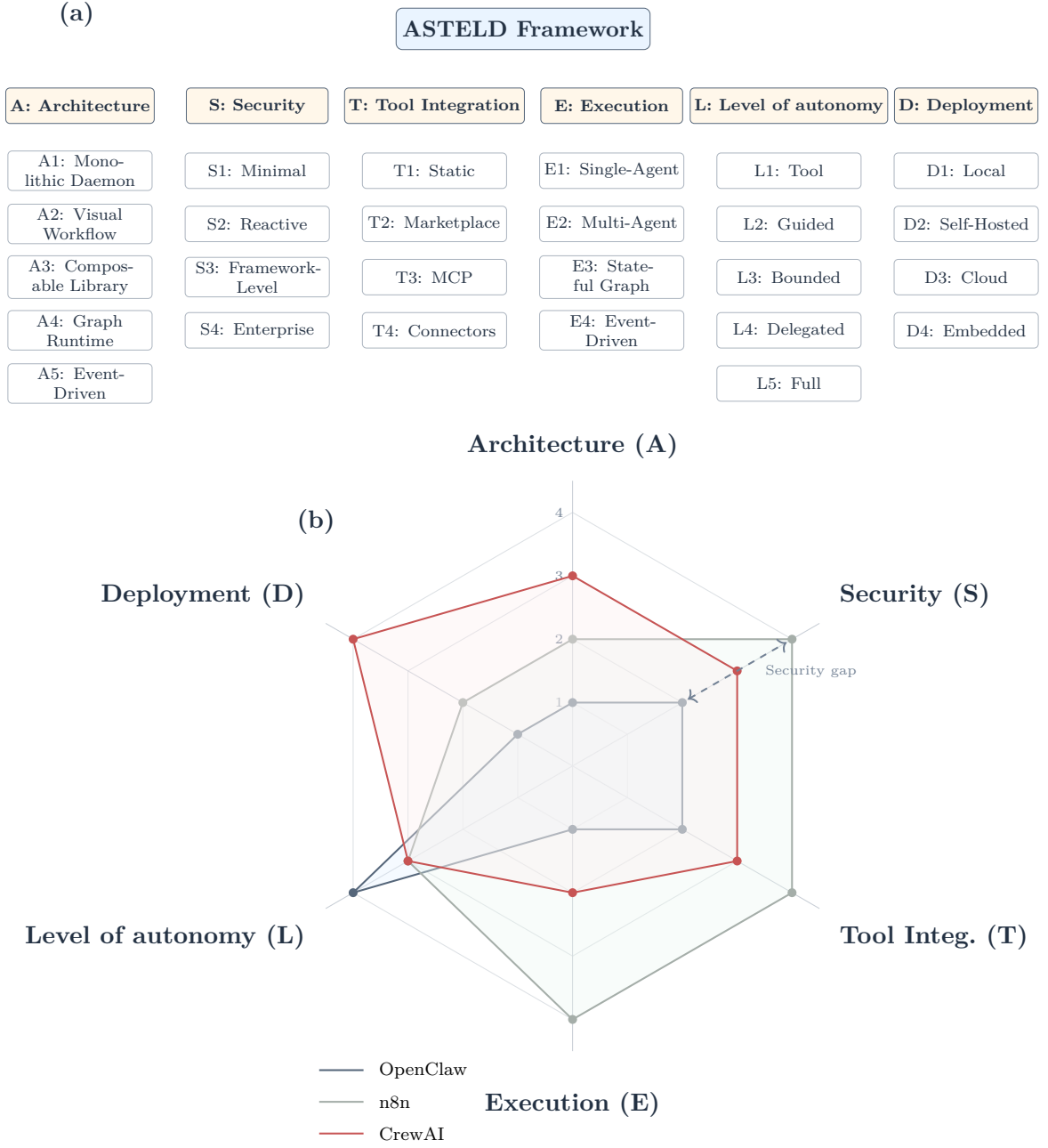

\subsection{Framework Mapping}

Table~\ref{tab:asteld} presents the complete ASTELD mapping for all eight frameworks. Under the dominant-configuration rule, the eight systems receive distinct six-axis profiles, satisfying the framework's basic discrimination criterion for this evaluation set.

\begin{table}[tbp]
\centering
\caption{ASTELD classification of eight autonomous AI agent platforms.}
\label{tab:asteld}
\footnotesize
\setlength{\tabcolsep}{4pt}
\begin{tabularx}{\textwidth}{@{}>{\RaggedRight\arraybackslash}p{2.10cm} *{6}{>{\centering\arraybackslash}p{0.95cm}} >{\centering\arraybackslash}p{1.05cm} Y@{}}
\toprule
\textbf{Framework} & \textbf{A} & \textbf{S} & \textbf{T} & \textbf{E} & \textbf{L} & \textbf{D} & \textbf{Stars} & \textbf{Primary Use Case} \\
\midrule
OpenClaw            & A1 & S2 & T2 & E1 & L2/\linebreak L4$^\dag$ & D1 & 360K+ & Personal AI assistant \\
Hermes             & A1 & S3 & T3 & E1 & L3/\linebreak L4 & D2 & 117K+ & Server-resident personal agent \\
AutoGPT             & A2 & S2 & T2 & E1$\to$\linebreak E4 & L4 & D2 & 180K+ & Autonomous research \\
CrewAI              & A3 & S3 & T3 & E2 & L3 & D4 & 50K+ & Multi-agent collaboration \\
LangGraph           & A4 & S3 & T3 & E3 & L2--L3 & D3/\linebreak D4 & 30K+ & Stateful workflows \\
n8n                 & A2 & S4 & T4 & E4 & L3 & D2/\linebreak D3 & 180K+ & Workflow automation \\
Dify                & A2 & S4 & T3 & E4 & L3 & D2/\linebreak D3 & 130K+ & Production AI apps \\
AutoGen / MS AF     & A5 & S3 & T4 & E2 & L3--L4 & D4 & 50K+ & Research collaboration \\
\bottomrule
\end{tabularx}
\vspace{2pt}

\raggedright\footnotesize
$^\dag$ OpenClaw is designed for L2 (per-action approval), but many users configure ``allow always,'' effectively operating at L4 (delegated autonomy)~\cite{microsoft2026running_openclaw}. Axis definitions: A = Architecture Pattern, S = Security Posture, T = Tool Integration Model, E = Execution Paradigm, L = Level of autonomy, D = Deployment Topology. Star counts were refreshed from GitHub repository pages in late April~2026~\cite{github_openclaw}\cite{github_hermes}\cite{github_autogpt}\cite{github_crewai}\cite{github_langgraph}\cite{github_n8n}\cite{github_dify}\cite{github_autogen}.
\end{table}

We acknowledge that some frameworks span multiple categories depending on configuration and usage context; assignments reflect their primary design and typical deployment.

OpenClaw's level of autonomy carries a notable caveat: the system is designed for L2 (guided autonomy with per-action approval), but the majority of users configure ``allow always'' mode for convenience, effectively operating at L4 (delegated autonomy) with an S2 (reactive) security posture~\cite{microsoft2026running_openclaw}. Hermes shows a related but distinct pattern: it remains architecturally monolithic (A1/E1) yet pushes toward a more server-resident and security-hardened S3/D2 profile through containerized terminal backends, persistent memory, and protocol-based extensibility~\cite{github_hermes}. These classifications reflect typical usage patterns rather than strictly enforced system constraints.

\subsection{Cross-Cutting Patterns}

The ASTELD mapping suggests three structural patterns that extend beyond individual frameworks, based on the configurations summarized in Table~\ref{tab:asteld}.

\textbf{Pattern~1: The Security-Accessibility Diagonal.} Plotting Security (S) against Deployment (D) across the surveyed frameworks, a general inverse trend can be observed. D1 (local-first) platforms tend to align with S1--S2 (minimal to reactive security), with OpenClaw as a representative example. D2/D3 (server/cloud) platforms are more often associated with S3--S4 (framework to enterprise-level security), as seen in n8n and Dify. D4 (library-embedded) platforms typically correspond to S3 (framework-level security), where responsibility is partially delegated to developers.

Notably, none of the surveyed frameworks occupy the combination of D1 + S4 (local-first deployment with enterprise-grade security). While this observation is limited to the current sample, it highlights a potentially underexplored design space. In practice, users seeking locally running personal assistants may encounter more limited security features, whereas systems offering stronger security guarantees are more commonly deployed in server- or cloud-based settings. This trade-off appears in the current ecosystem, though further evidence would be needed to determine whether it reflects a fundamental constraint or a transient design choice.

\textbf{Pattern~2: Execution-Architecture Coupling.} Architecture pattern appears closely associated with execution paradigm. For example, A1 (monolithic) systems are typically paired with E1 (single-agent loops), A2 (visual workflow) with E4 (event-driven pipelines), A3 (composable libraries) with E2 (multi-agent interaction), A4 (graph runtimes) with E3 (stateful graph execution), and A5 (event-driven multi-agent systems) also with E2. This alignment suggests that architectural design choices often influence, or co-evolve with, execution models. However, this relationship is empirical rather than strictly deterministic, and alternative pairings may be possible.

\textbf{Pattern~3: Convergence with Persistent Differentiation.} Several frameworks exhibit convergence toward a common set of capabilities, including model-agnostic design, support for tool integration protocols (e.g., MCP), visual workflow construction, and human-in-the-loop control mechanisms. At the same time, meaningful architectural differences remain. The ecosystem can be broadly interpreted as forming three functional groupings: personal assistant-oriented systems (e.g., OpenClaw, Hermes), multi-agent orchestration frameworks (e.g., CrewAI, LangGraph, AutoGen), and workflow automation platforms (e.g., n8n, Dify). While these groupings are not mutually exclusive, they reflect typical usage patterns observed in the current landscape. Cross-group competition appears more limited than within-group variation, although this characterization is based on a small set of representative systems~\cite{letsdatascience2026langgraph}\cite{github_hermes}.

\subsection{Comparative Analysis by Dimension}

\textbf{Architecture.} OpenClaw's monolithic daemon (A1) is distinctive in using messaging platforms as its primary user interface, which may lower the barrier to entry for non-technical users but can introduce additional challenges for enforcing security boundaries. Hermes occupies a nearby architectural niche but relocates the agent to a server-resident D2 deployment with persistent memory, MCP extensibility, and a built-in learning loop, making it a more direct competitor in the personal-agent lane than the workflow-centric or library-centric alternatives~\cite{github_hermes}. AutoGPT, n8n, and Dify share the visual workflow pattern (A2), although their underlying system designs differ; notably, n8n's node-based canvas predates AI agent frameworks and originates from general workflow automation~\cite{n8n2026platform}. LangGraph's graph-based runtime (A4) enables expressive control flow, including cycles and stateful execution, but may require greater familiarity with graph-based abstractions~\cite{langchain2025langgraph}.

\textbf{Security.} The surveyed frameworks exhibit a range of security capabilities. n8n and Dify provide features such as RBAC, SSO/SAML, audit logging, and support for controlled deployment environments, reflecting their positioning toward enterprise use cases~\cite{n8n2026platform}\cite{dify2025milestone}. In contrast, OpenClaw is characterized here as S2 (reactive), relying more heavily on post-hoc controls and user configuration. While OpenClaw has a large deployment footprint, this comparison highlights a potential gap between deployment scale and the maturity of built-in security mechanisms. This observation is based on available feature sets and reported configurations rather than a formal security evaluation.

\textbf{Tool Integration.} The Model Context Protocol (MCP) is increasingly adopted across frameworks, with several platforms supporting protocol-based tool integration~\cite{langchain2025langgraph}\cite{n8n2026platform}\cite{dify2025milestone}. OpenClaw follows a T2 (marketplace-based) approach, enabling rapid expansion of available tools. Prior work has noted that marketplace-based ecosystems may introduce risks if contributions are not systematically vetted~\cite{cyberpress2026clawhavoc}\cite{koi2026clawhavoc}. The observed shift toward protocol-based integration (T3) in some systems may reflect an effort to standardize and better control tool invocation, although both models continue to coexist in practice.

\textbf{Human-in-the-Loop.} Human-in-the-loop (HITL) mechanisms vary in granularity and integration. LangGraph provides support for checkpointing, state inspection, and intervention at arbitrary points in execution graphs~\cite{langchain2025langgraph}. n8n supports human intervention through workflow nodes, enabling approval steps and user input routing within pipelines~\cite{n8n2026platform}. OpenClaw includes an approval-based mechanism intended to gate actions, although its effectiveness depends on user configuration and system context (see Section~\ref{sec:security}). Overall, HITL support is present across frameworks but differs in depth and flexibility.

\textbf{License and Governance.} The surveyed frameworks adopt a range of licensing models. OpenClaw, CrewAI, LangGraph, and AutoGen use permissive MIT licenses, while Dify adopts Apache~2.0. AutoGPT applies a Polyform Shield License to parts of its platform, and n8n uses a fair-code model (Sustainable Use License) alongside commercial offerings. These choices reflect differing approaches to ecosystem growth, monetization, and long-term maintenance. While permissive licenses may facilitate adoption, they do not inherently provide mechanisms for coordinated governance or resource allocation, which may influence sustainability considerations.

\subsection{Framework Selection Matrix}

Based on the comparative analysis, we propose a selection matrix for practitioners:

\begin{table}[tbp]
\centering
\caption{Agent-platform selection matrix for practitioners.}
\label{tab:selection_matrix}
\footnotesize
\setlength{\tabcolsep}{4pt}
\begin{tabularx}{\textwidth}{@{}>{\RaggedRight\arraybackslash}p{3.15cm} >{\RaggedRight\arraybackslash}p{2.65cm} Y@{}}
\toprule
\textbf{Use Case} & \textbf{Recommended} & \textbf{Rationale} \\
\midrule
Personal AI via messaging & OpenClaw & Only native WhatsApp/Telegram/Discord integration \\
Multi-agent collaboration & CrewAI & Purpose-built for role-based orchestration; 5.76x faster than LangGraph~\cite{letsdatascience2026langgraph} \\
Complex stateful workflows & LangGraph & Most expressive graph control flow; best HITL; checkpoint/resume~\cite{langchain2025langgraph} \\
Visual no-code agents & n8n or Dify & 400--600+ integrations; enterprise RBAC/SSO~\cite{n8n2026platform}\cite{dify2025milestone} \\
Autonomous research & AutoGPT & Pioneer in autonomous task loops; visual workflow builder~\cite{github_autogpt} \\
Enterprise multi-agent & MS Agent Framework & Azure integration; 1,000+ connectors via Semantic Kernel~\cite{microsoftlearn2026agent_framework} \\
Maximum security & n8n (self-hosted) or TrustClaw & RBAC/SSO/audit/SIEM; or OpenClaw fork with OAuth + sandbox~\cite{n8n2026platform}\cite{ossinsight2026forks_wave} \\
\bottomrule
\end{tabularx}
\end{table}

\subsection{Summary}

The evaluation supports three uses of ASTELD. First, it discriminates among the eight platforms through distinct dominant profiles. Second, it exposes structural relationships that are difficult to state consistently in narrative comparisons, including the security--accessibility diagonal and execution--architecture coupling. Third, it identifies design gaps, most notably the absence of a D1 + S4 platform in the evaluation set. The framework does not impose a universal ranking: OpenClaw, n8n, and LangGraph optimize different coordinates for different use cases. The sample remains limited and assignments reflect dominant configurations rather than every optional deployment. The derivative ecosystem examined next provides a separate explanatory test: if ASTELD captures consequential constraints, derivative projects should modify the axes on which the base system is most restricted.

\section{Adoption, Social Impact, and the Derivative Ecosystem}
\label{sec:adoption}

OpenClaw's adoption trajectory is unprecedented in open-source history, both in its growth and in the breadth of its social impact. This section examines three interrelated phenomena: quantitative growth dynamics, the sociotechnical implications of mass adoption, and the derivative ecosystem that emerged as a response to OpenClaw's architectural constraints. The derivative data serve as an explanatory evaluation of ASTELD: they allow us to test whether ecosystem innovation is concentrated on the axes where the base platform exhibits the clearest limitations.

\begin{figure}[tbp]
\centering
\definecolor{ink}{HTML}{223042}
\definecolor{line}{HTML}{55667A}
\definecolor{muted}{HTML}{738295}
\definecolor{panel}{HTML}{F7F9FC}
\definecolor{accent}{HTML}{C75454}
\begin{tikzpicture}
\begin{axis}[
    width=0.98\columnwidth,
    height=6.5cm,
    xlabel={Date},
    ylabel={GitHub Stars},
    xmin=0, xmax=155,
    ymin=0, ymax=380,
    grid=major,
    grid style={muted!20},
    tick label style={font=\small, text=ink},
    label style={font=\small, text=ink},
    xtick={7,38,69,98,128},
    xticklabels={December, 2026, February, March, April},
    ytick={0,100,200,300},
    yticklabels={0,100K,200K,300K},
    legend style={
        font=\scriptsize,
        at={(0.02,0.98)},
        anchor=north west,
        draw=line,
        fill=panel,
        fill opacity=0.8,
        text opacity=1,
        rounded corners=2pt
    },
]

\addplot[
    ultra thick, accent, smooth
] coordinates {
    (0,0)
    (10,0.3)
    (20,0.8)
    (30,1.5)
    (40,2.8)
    (50,4.2)
    (58,5.4)
    (62,6.8)
    (65,8.5)
    (68,10.5)
    (69,40)
    (70,60)
    (73,88)
    (76,132)
    (80,190)
    (83,202)
    (88,215)
    (92,224)
    (96,244)
    (99,250.829)
    (104,272)
    (110,292)
    (116,310)
    (120,325)
    (125,338)
    (132,349)
    (140,356.8)
    (152,360.2)
};
\addlegendentry{OpenClaw}

\node[font=\scriptsize, anchor=south west, text=accent] at (axis cs:70, 64) {60K (Feb.~2)};
\node[font=\scriptsize, anchor=south west, text=accent] at (axis cs:92, 230) {$>$ Linux (Feb.~24)};
\node[font=\scriptsize, anchor=south west, text=accent] at (axis cs:99, 258) {250.8K (Mar.~3)};
\node[font=\scriptsize, anchor=south west, text=accent] at (axis cs:120, 332) {335K (Mar.~24)};
\node[font=\scriptsize, anchor=south west, text=accent] at (axis cs:152, 366) {360K+ (late Apr.)};

\draw[dashed, muted] (axis cs:0,243.438) -- (axis cs:99,243.438);
\node[font=\scriptsize, text=muted, anchor=south east] at (axis cs:95, 243.438) {React reference};
\draw[dashed, muted] (axis cs:0,218) -- (axis cs:92,218);
\node[font=\scriptsize, text=muted, anchor=south east] at (axis cs:88, 218) {Linux reference};

\end{axis}
\end{tikzpicture}
\caption{OpenClaw GitHub star trajectory with cross-checked milestone anchors. Dashed horizontal lines mark the Linux and React comparison levels used in the text.}
\label{fig:growth}
\end{figure}
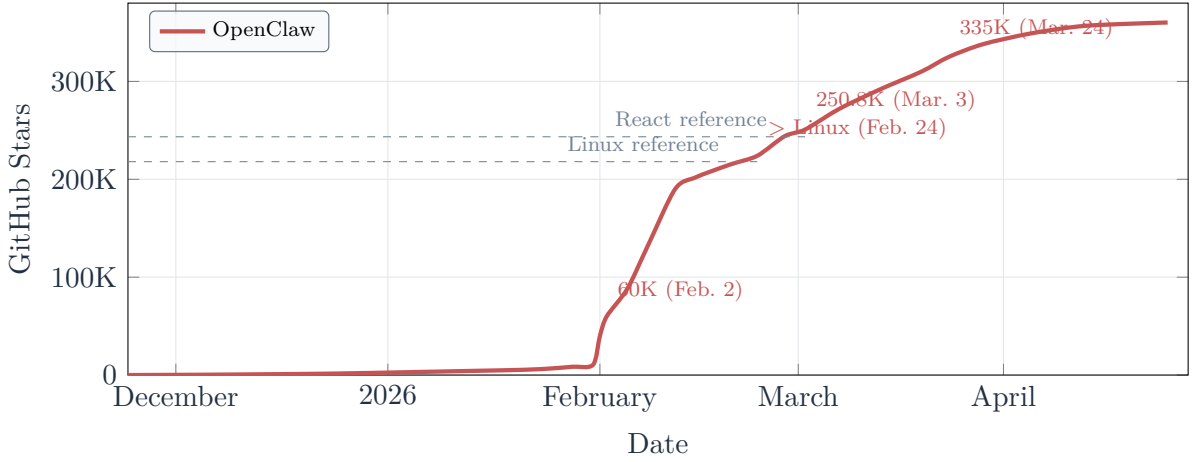

\subsection{Adoption Metrics and Growth Dynamics}

\subsubsection{GitHub Star Growth}

Figure~\ref{fig:growth} shows a near-flat pre-viral baseline through December and most of January, a sharp inflection beginning on January~29, sustained high-velocity growth through mid-February, and a later March reacceleration. In terms of the phase definitions introduced in Section~\ref{sec:origins}, the initial growth wave spans Phase~1 (January~29 -- February~2) and the early portion of Phase~2, while the March reacceleration corresponds to Phase~3.

The velocity comparison underscores the anomaly. OpenClaw reached 100,000~stars in 12~days; React required approximately 10~years (68~stars/day average); the Linux kernel required approximately 30~years (18~stars/day); AutoGPT, the previous fastest-growing AI project, required approximately 6~months~\cite{ossinsight2026forks_wave}\cite{starhistory2024ai_agents}\cite{starhistory2026openclaw}. OpenClaw's 100K-star velocity is 123~times that of React and 419~times that of Linux.

The March reacceleration is closely tied to geographic adoption dynamics. Traffic from China surged 1,436\% month-over-month during March~2026, propelled by what Chinese media termed the ``lobster craze''~\cite{gradually2026statistics}\cite{starhistory2026openclaw}. This China-driven growth phase was accompanied by a surge in Chinese-language tutorials and enterprise experimentation (including Tencent's ClawPro).

\subsubsection{Comparative Platform Metrics}

The geographic distribution of traffic reveals concentrated adoption in three regions: the United States (16.29\%), India (12.16\%), and China (12.08\%, with month-over-month growth of 1,436\%), followed by Germany (4.10\%, +992\% MoM) and Canada (3.53\%, +1,259\% MoM)~\cite{starhistory2026openclaw}\cite{fatjoe2026openclaw_stats}.

\subsubsection{Usage and Ecosystem Metrics}

Beyond GitHub metrics, OpenClaw's adoption is characterized by the following indicators as of April~2026: 3M+ monthly active users; 30M+ website visitors per month; 500,000+ running instances across 80+ countries; approximately 500,000+ internet-facing instances in late-March~2026 live checks, of which 15,000+ were reported as exploitable via known RCE flaws; 40,000+ skills on ClawHub; 1,800+ repository contributors; 2M+ total skill installations; tokens processed via 350+ LLM models; 180+ ecosystem startups with combined monthly revenue exceeding \$320,000; and 80+ npm packages dependent on the OpenClaw runtime~\cite{gradually2026statistics}\cite{openclawvps2026statistics}\cite{securityscorecard2026attack_surface}\cite{columbus2026instances}\cite{github_openclaw}.

\subsubsection{\texorpdfstring{The Adoption Paradox: Stars $\neq$ Enterprise Adoption}{The Adoption Paradox: Stars != Enterprise Adoption}}

A critical finding is the disconnect between community popularity and enterprise adoption. LangGraph, with 30,000+~stars, is deployed by 400~companies in production---including Cisco, Uber, LinkedIn, BlackRock, and JPMorgan Chase---and registers 34.5~million monthly package downloads~\cite{langchain2025langgraph}\cite{letsdatascience2026langgraph}\cite{github_langgraph}. Dify, with 130,000+~stars, serves 280~enterprises with 1.4~million deployments and a \$30~million fundraise~\cite{dify2025milestone}\cite{github_dify}. OpenClaw, with 360,000+~stars---more than LangGraph and Dify combined---has 180+ ecosystem startups, \$320,000 per month in ecosystem revenue, and zero Fortune~500 customers~\cite{gradually2026statistics}\cite{starhistory2026openclaw}\cite{openclawvps2026statistics}\cite{github_openclaw}.

This paradox illustrates a systematic relationship: GitHub stars correlate with accessibility and viral potential, not with production maturity or enterprise trust. Enterprise adopters require security guarantees (RBAC, SSO, audit trails), operational reliability (SLAs, support contracts), and regulatory compliance---none of which OpenClaw provides. The ASTELD framework captures this relationship: OpenClaw's S2 security posture is incompatible with enterprise requirements, regardless of its D1 deployment accessibility or its community scale.

\subsection{The Derivative Ecosystem}

The most striking structural phenomenon in OpenClaw's ecosystem is the rapid emergence of derivative projects: over 50~forks, rewrites, and inspired-by projects appeared within three months of OpenClaw's breakout~\cite{ossinsight2026forks_wave}. This fragmentation is not random; it is a systematic response to specific architectural constraints that the ASTELD framework makes legible.

\subsubsection{Derivative Taxonomy}

We classify derivatives into three tiers based on their relationship to the OpenClaw codebase and their target audience.

\begin{table}[tbp]
\centering
\caption{OpenClaw derivative ecosystem classification.}
\label{tab:derivatives}
\footnotesize
\setlength{\tabcolsep}{4pt}
\newcolumntype{C}{>{\RaggedRight\arraybackslash}p{1.82cm}}
\newcolumntype{D}{>{\centering\arraybackslash}p{1.45cm}}
\newcolumntype{E}{>{\centering\arraybackslash}p{1.30cm}}
\begin{tabularx}{\textwidth}{@{}>{\RaggedRight\arraybackslash}p{1.85cm} C C D E E Y@{}}
\toprule
\textbf{Project} & \textbf{Tier} & \textbf{Lang.} & \textbf{Created} & \textbf{Stars} & \textbf{Forks} & \textbf{ASTELD / Key Differentiation} \\
\midrule
\multicolumn{7}{@{}l}{\textit{Tier 1: Community Derivatives}} \\
\addlinespace
Nanobot (HKUDS)  & Commu\-nity & Python      & 2026-02-01 & 37.4K+ & 6.5K+ & Reduce complexity; compact Python rewrite \\
ZeroClaw         & Commu\-nity & Rust        & 2026-02-13 & 30.6K+ & 4.5K+ & S2$\to$S3 (memory safety); 8.8\,MB binary \\
PicoClaw (Sipeed)& Commu\-nity & Go          & 2026-02-04 & 28.5K+ & 4.1K+ & D1 with edge/IoT emphasis \\
NanoClaw         & Commu\-nity & TypeScript  & 2026-01-31 & 28K+   & 12.6K+ & Simplify A1 (700 LoC); fork ratio 0.4+ \\
TrustClaw        & Commu\-nity & TypeScript  & 2026-02    & Private & Private & S2$\to$S3 (OAuth + sandbox) \\
\addlinespace
\multicolumn{7}{@{}l}{\textit{Tier 2: Enterprise Adaptations}} \\
\addlinespace
Tencent ClawPro  & Enterprise & ---        & 2026-03    & Private & Private & A1$\to$A2, S2$\to$S4, D1$\to$D2 \\
QClaw            & Enterprise & ---        & 2026-03    & Private & Private & WeChat integration; China LLM providers \\
MaxClaw (Minimax)& Enterprise & ---        & 2026-03    & Private & Private & China market optimization \\
Kimi Claw        & Enterprise & ---        & 2026-03    & Private & Private & Moonshot AI integration \\
\addlinespace
\multicolumn{7}{@{}l}{\textit{Tier 3: Research Platforms}} \\
\addlinespace
Molt\-worker    & Research & TypeScript   & 2026-01    & 9.8K+ & 1.8K+ & A1$\to$A5, E1$\to$E2 (Cloudflare sandbox deployment) \\
Dr.~Claw         & Research & TypeScript   & 2026-01    & 900+ & 90+ & A1$\to$A2, E1$\to$E4 (research pipeline) \\
\bottomrule
\end{tabularx}
\vspace{2pt}

\raggedright\footnotesize
Combined Tier~1 metrics (top 4): roughly 124.5K+ stars and 27.7K+ forks within a 14-day emergence window (Jan~31 -- Feb~13, 2026). Community-fork metrics were refreshed from current GitHub repository pages in late April~2026~\cite{github_nanobot}\cite{github_zeroclaw}\cite{github_picoclaw}\cite{github_nanoclaw}. Closed-source derivatives are marked as Private; the final-column notes are derived from documentation and feature analysis.
\end{table}

\textbf{Tier~1: Community Derivatives.} Rather than direct modifications of the OpenClaw codebase, the most prominent early derivatives were fresh implementations that re-expressed OpenClaw's core assistant model in Python, Rust, Go, and TypeScript for different priorities of simplicity, hardware efficiency, portability, and customizability~\cite{ossinsight2026forks_wave}. NanoClaw's current fork ratio is notably anomalous at approximately 0.4+. Combined, the top four community derivatives accumulated roughly 124,500+~stars and 27,700+~forks by late April~2026~\cite{github_nanobot}\cite{github_zeroclaw}\cite{github_picoclaw}\cite{github_nanoclaw}. TrustClaw, while community-developed, specifically targets security hardening by adding OAuth and sandboxed execution (S2$\to$S3).

\textbf{Tier~2: Enterprise Adaptations.} Commercial or institutional projects that extend OpenClaw for enterprise requirements. Tencent's ClawPro shifts from A1 to A2 (visual workflow) and S2 to S4 (enterprise RBAC/SSO), targeting Chinese enterprise customers. These projects uniformly shift the Security (S) and Deployment (D) axes toward enterprise-grade values, confirming that OpenClaw's S2/D1 profile is structurally incompatible with enterprise requirements.

\textbf{Tier~3: Research Platforms.} Projects that use OpenClaw as a foundation for academic research rather than production deployment. Moltworker shifts from A1 to A5 (multi-agent) and E1 to E2 (multi-agent conversation), enabling research on agent collaboration and delegation. Dr.~Claw shifts E1 to E4 (research pipeline workflow), enabling structured experimental workflows.

\begin{figure}[tbp]
\centering
\definecolor{ink}{HTML}{223042}
\definecolor{line}{HTML}{55667A}
\definecolor{muted}{HTML}{738295}
\definecolor{iface}{HTML}{EAF4FF}
\definecolor{control}{HTML}{ECF8F1}
\definecolor{exec}{HTML}{F3EEFF}
\definecolor{riskfill}{HTML}{FDEEEF}
\definecolor{risk}{HTML}{C75454}
\begin{tikzpicture}[
    >=Stealth,
    node distance=0.8cm,
    base/.style={
        circle,
        draw=line,
        fill=iface,
        minimum size=0.92cm,
        font=\scriptsize\bfseries, text=ink,
        inner sep=1pt
    },
    fork/.style={
        circle,
        draw=control!70!black,
        fill=control,
        minimum size=0.72cm,
        font=\tiny\bfseries, text=ink,
        inner sep=1pt
    },
    ent/.style={
        circle,
        draw=muted,
        fill=exec,
        minimum size=0.72cm,
        font=\tiny\bfseries, text=ink,
        inner sep=1pt
    },
    res/.style={
        circle,
        draw=risk,
        fill=riskfill,
        minimum size=0.72cm,
        font=\tiny\bfseries, text=ink,
        inner sep=1pt
    },
    axlabel/.style={
        font=\scriptsize\bfseries,
        text=muted
    },
    ticklabel/.style={
        font=\tiny,
        text=muted
    },
    shiftvec/.style={
        ->,
        thick,
        draw=muted
    },
    blabel/.style={
        font=\scriptsize,
        text=line,
        align=center
    },
    glabel/.style={
        font=\tiny,
        text=control!70!black,
        align=left
    },
    olabel/.style={
        font=\tiny,
        text=muted,
        align=left
    },
    plabel/.style={
        font=\tiny,
        text=risk,
        align=left
    },
    legend/.style={
        font=\tiny,
        anchor=north west,
        align=left
    }
]

\draw[->, thick, muted!40] (-0.50, 0.00) -- (10.20, 0.00);
\node[axlabel, anchor=west] at (9.15, -0.28) {Security (S) $\longrightarrow$};

\foreach \x/\lbl in {0.00/S1, 2.50/S2, 5.00/S3, 7.50/S4} {
    \draw[muted!40] (\x, -0.13) -- (\x, 0.13);
    \node[ticklabel, below] at (\x, -0.13) {\lbl};
}

\draw[->, thick, muted!40] (0.00, -0.55) -- (0.00, 7.90);
\node[axlabel, anchor=south, rotate=90] at (-0.28, 6.20) {Execution (E) $\longrightarrow$};

\foreach \y/\lbl in {0.00/E1, 2.50/E2, 5.00/E3, 7.50/E4} {
    \draw[muted!40] (-0.13, \y) -- (0.13, \y);
    \node[ticklabel, left] at (-0.16, \y) {\lbl};
}

\node[base] (oc) at (2.50, 0.00) {OC};
\node[blabel, anchor=north] at (2.50, -0.72) {OpenClaw\\360K+ $\bigstar$};

\node[fork] (pc) at (0.85, 0.35) {PC};
\node[glabel, anchor=east, align=right] at (0.55, 0.35) {PicoClaw\\(edge)};

\node[fork] (nb) at (1.55, 1.15) {NB};
\node[glabel, anchor=south, align=center] at (1.55, 1.55) {Nanobot\\(30K LoC)};

\node[fork] (nc) at (2.55, 0.85) {NC};
\node[glabel, anchor=west] at (3.15, 0.88) {NanoClaw (700 LoC)};

\node[fork] (zc) at (5.00, 0.45) {ZC};
\node[glabel, anchor=west] at ([xshift=4pt]zc.east) {ZeroClaw\\(Rust)};

\node[fork] (tc) at (5.00, -0.42) {TC};
\node[glabel, anchor=north, align=center] at (5.00, -0.83) {TrustClaw (OAuth)};

\node[ent] (qc) at (5.50, 1.80) {QC};
\node[olabel, anchor=west] at (6.12, 1.80) {QClaw};

\node[ent] (cp) at (7.50, 7.00) {CP};
\node[olabel, anchor=south, align=center] at (7.50, 7.43) {Tencent ClawPro};

\node[res] (mw) at (2.50, 2.50) {MW};
\node[plabel, anchor=east, align=right] at (1.95, 2.50) {Moltworker};

\node[res] (dc) at (5.00, 7.00) {DC};
\node[plabel, anchor=south, align=center] at (5.00, 7.43) {Dr.\ Claw};

\draw[shiftvec] (oc.east) -- (zc.west);
\draw[shiftvec] ([yshift=-2pt]oc.east) -- (tc.west);
\draw[shiftvec] ([xshift=1pt,yshift=2pt]oc.north east) -- (qc.south west);
\draw[shiftvec] (oc.north) -- (mw.south);
\draw[shiftvec] (oc.north east) .. controls (3.70, 3.80) and (4.55, 5.60) .. (dc.south);
\draw[shiftvec] (oc.north east) .. controls (4.25, 2.00) and (6.00, 5.30) .. (cp.south west);

\draw[shiftvec] (oc.west) -- (pc.east);
\draw[shiftvec] (oc.north west) -- (nb.south east);
\draw[shiftvec] (oc.north) -- (nc.south);

\draw[dashed, risk, thick, rounded corners=3pt] (6.45, 4.55) rectangle (9.55, 6.80);
\node[font=\tiny, text=risk, align=center] at (8.00, 5.68) {Empty quadrant:\\[1pt]
D1 + S4 + E3/E4\\[1pt]
(unsolved design\\challenge)};

\node[legend, text width=3.3cm] at (6.65, -1.15) {
    \textcolor{line}{$\bullet$} OpenClaw (base)\\
    \textcolor{control!70!black}{$\bullet$} Tier 1: Community forks\\
    \textcolor{muted}{$\bullet$} Tier 2: Enterprise adaptations\\
    \textcolor{risk}{$\bullet$} Tier 3: Research platforms\\
    $\longrightarrow$ ASTELD shift vector
};

\end{tikzpicture}

\caption{OpenClaw derivative ecosystem on the ASTELD Security--Execution plane. Arrows indicate the shift from OpenClaw's base position (S2, E1).}
\label{fig:derivative_map}
\end{figure}
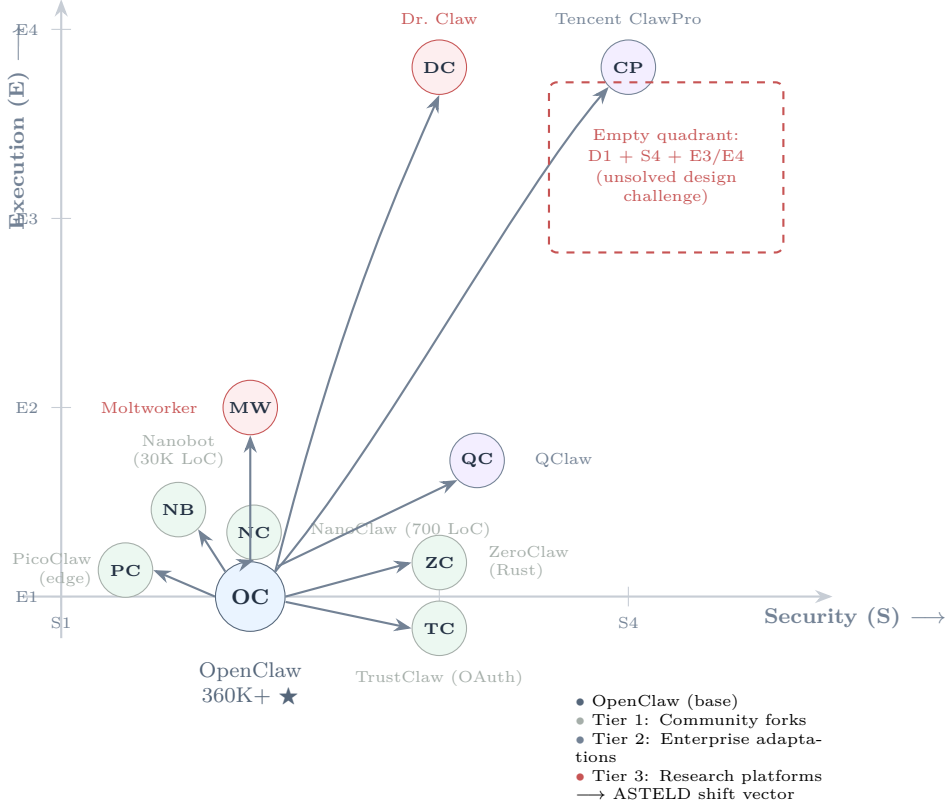

\subsubsection{Derivative Prediction Power}

The ASTELD framework reveals that derivative shifts are non-random. Three axes---Security (S), Execution (E), and Deployment (D)---account for the vast majority of derivative innovation. No derivative has shifted the Tool Integration axis from T2 to T3/T4 (marketplace to protocol-based), suggesting that the marketplace model is not perceived as the primary constraint by derivative developers, despite its role in ClawHavoc. Similarly, no derivative has shifted the Architecture axis from A1 to A4 (monolithic to graph-based), suggesting that the monolithic daemon pattern is perceived as a feature, not a bug, by the community.

This pattern has predictive implications. Future derivatives are most likely to target the Security-Deployment gap identified in Section~\ref{sec:ecosystem}: a local-first agent with enterprise-grade security. The current empty quadrant (D1 + S4) represents the most commercially valuable unsolved design challenge in the autonomous agent space.

\subsection{Social and Regulatory Impact}

\subsubsection{The Token Economy}

OpenClaw has catalyzed a nascent token economy. As of April~2026, OpenClaw instances are estimated to have collectively processed trillions of tokens across 350+ language models~\cite{gradually2026statistics}. The 180+ ecosystem startups---skill developers, hosting providers, integration services---generate combined monthly revenue exceeding \$320,000~\cite{gradually2026statistics}\cite{starhistory2026openclaw}\cite{openclawvps2026statistics}. While modest relative to the broader AI industry (\$7.84~billion global AI agent market in 2025, projected to reach \$52.62~billion by 2030 at 46.3\% CAGR~\cite{starhistory2026openclaw}\cite{columbus2026instances}), this ecosystem represents the first organic token-mediated economy around an open-source AI agent.

\subsubsection{ClawHub Marketplace Dynamics}

The ClawHub marketplace has grown to 40,000+ skills. However, the security profile is concerning: independent assessment finds 47\% of skills assessed as safe, 36\% containing prompt injection vectors, 8\% attempting data exfiltration, and 6\% requesting excessive permissions~\cite{gradually2026statistics}. The trend is worsening: at the time of the ClawHavoc disclosure in February~2026, approximately 12\% of the then-10,700 skills were classified as malicious~\cite{cyberpress2026clawhavoc}; by April, 36\% of the now-40,000+ skills contain prompt injection~\cite{gradually2026statistics}\cite{openclawvps2026statistics}.

Because these two assessments use different classification schemes---broader unsafe or adversarial characteristics in the former case, versus explicitly malicious skills in the latter---they are not directly comparable as a single time-series. Even so, taken together, they indicate that marketplace risk is worsening faster than marketplace governance is maturing. Therefore, the governance challenge---enabling permissionless contribution while preventing compromise---remains the central unsolved problem for agent skill ecosystems (see RQ5, Section~\ref{sec:rq}).

\subsubsection{Regulatory Responses}

OpenClaw has begun to draw governmental attention to autonomous AI agents, though no formal multi-jurisdictional review process has yet been publicly confirmed.

The broader regulatory landscape remains uncertain. The EU AI Act classifies AI systems by risk level but does not specifically address autonomous agents with system-level access. The NIST AI Risk Management Framework provides general principles but no agent-specific guidance. No academic analysis has mapped how existing regulatory frameworks apply to the specific capabilities and risks of autonomous AI agents (see RQ7, Section~\ref{sec:rq}).

\subsubsection{Enterprise Adoption Dynamics}

The enterprise adoption paradox identified in Section~7.1.4 reflects a broader market dynamic. Gartner projects that 40\% of enterprise applications will incorporate AI agents by end of~2026 (up from less than 5\% in 2025), and 80\% of Fortune~500 companies are exploring agent deployments~\cite{starhistory2026openclaw}. Yet this adoption is overwhelmingly concentrated in enterprise-grade frameworks: LangGraph (400 production deployments among named enterprises), Dify (280~enterprises, \$30M fundraise), and Microsoft Agent Framework (Azure integration, Semantic Kernel ecosystem)~\cite{langchain2025langgraph}\cite{dify2025milestone}\cite{microsoftlearn2026agent_framework}.

OpenClaw's path to enterprise adoption is blocked by structural factors captured in its ASTELD profile: S2 security posture (no RBAC, no SSO, no audit trails), D1 deployment topology (no server-grade deployment option by default), and the ongoing security advisory accumulation (averaging more than one per day)~\cite{jgamblin2026openclawcves}. The derivative ecosystem's enterprise tier (Tencent ClawPro, QClaw) represents market attempts to bridge this gap, but none has yet achieved the scale of LangGraph or Dify in enterprise settings.

\subsection{Summary}

OpenClaw's adoption metrics are record-setting by every community measure: stars, growth velocity, monthly active users, and ecosystem breadth. Yet the adoption paradox---360,000+ stars but zero Fortune~500 customers---reveals that community scale and enterprise readiness are orthogonal dimensions. The derivative ecosystem, with 50+ projects appearing in three months, is not fragmentation for its own sake but a systematic market response to specific ASTELD-axis constraints. The three most constrained axes---Security, Execution, and Deployment---predict the direction of derivative innovation with remarkable accuracy. The social and regulatory impacts---the token economy, ClawHub's worsening security profile, emerging regulatory attention, and the enterprise adoption gap---point to fundamental open questions that we formulate in the next section.

\section{Open Research Questions and Future Directions}
\label{sec:rq}

The preceding analysis reveals a field in rapid but uneven maturation. The technical capability of autonomous AI agents has advanced ahead of security infrastructure, governance frameworks, and academic analysis. This section formulates eight research questions grounded in the evidence gaps identified throughout this survey, organized by research domain.

\subsection{Architectural Research}

\textbf{RQ1: Can we design a formally verifiable architecture that enforces a strict separation between instruction channels and data channels in autonomous AI agents, thereby eliminating prompt injection as an authorization vulnerability?}

As shown in Section~\ref{sec:architecture}, OpenClaw adopts a monolithic architecture in which the LLM controller processes both user instructions and untrusted external data within a shared context window, without any privilege separation between reasoning and execution components. Section~\ref{sec:security} further demonstrates that this design directly enables multiple classes of prompt injection attacks, including indirect injection, guidance injection, and memory poisoning, all of which exploit the absence of a structural boundary between instruction and data processing.

These findings collectively suggest that prompt injection is not an isolated vulnerability, but a direct consequence of architectural design. Existing approaches, such as Energent.ai's multi-step parsing, attempt partial separation but lack formal verification or formal security guarantees~\cite{esecurityplanet2026prompt_injection}. This motivates the need for a provably secure architectural paradigm that enforces instruction--data separation at the system level. The relevant formalism may draw on information flow control, capability-based security, or language-theoretic security.

\textbf{RQ2: What is the principled design of a capability-based permission model for AI agent tool ecosystems that balances security, usability, and composability?}

As shown in Section~\ref{sec:architecture}, OpenClaw's tool execution engine operates with the full privileges of the user account, without any built-in mechanism for permission scoping or isolation. Section~\ref{sec:security} further reveals that this unrestricted access is a key enabler of multiple attack vectors, particularly supply chain attacks (Category~E) and prompt injection escalations (Category~D), where malicious skills or injected instructions can trigger arbitrary system-level actions.

The ClawHavoc campaign provides empirical evidence that the absence of a permission model allows large-scale compromise through seemingly benign skill installations. More broadly, these findings indicate that the current implicit full-trust model is fundamentally incompatible with secure agent deployment.

CertiK accordingly recommended that skills should declare resource needs upfront, analogous to Android's permission model~\cite{certik2026openclaw_report}. However, no existing framework implements such a mechanism at the architectural level. The design space includes static declarations at install time, dynamic capability negotiation at runtime, and sandboxed execution with permission escalation. A central challenge is to determine the appropriate permission granularity and enforcement model that can preserve open-source contribution velocity while reducing the risk of systemic compromise.

\subsection{Security Research}

\textbf{RQ3: How can the correctness of human-in-the-loop approval mechanisms in autonomous agents be formally defined and evaluated, given the observed gap between approval-time representations and execution-time behavior?}

As shown in Section~\ref{sec:security}, OpenClaw's approval system has been bypassed through multiple CVEs, each exploiting inconsistencies between the approval-time display of a command and its execution-time resolution. These cases reveal a systematic misalignment between what users approve and what the system ultimately executes.

This problem is analogous to time-of-check-to-time-of-use (TOCTOU) vulnerabilities, but arises at a higher abstraction level where the approved entity is a natural language or semi-structured representation of an action, rather than a fully specified program. A key open challenge is to define what constitutes ``approval correctness'' in such systems and to determine whether it can be formally specified, verified, or empirically evaluated under realistic agent behavior.

\textbf{RQ4: How can AI agent supply chain attacks be detected and mitigated when malicious behavior is encoded in natural language rather than executable code?}

Section~\ref{sec:security} demonstrates that agent supply chain attacks, such as the ClawHavoc campaign, operate through natural language payloads embedded in skill descriptions and execution flows, rather than traditional executable artifacts. As a result, conventional static analysis tools, including VirusTotal, are ineffective in identifying such threats~\cite{koi2026clawhavoc}\cite{arnav2026prompt_injection}.

These observations suggest a shift in the nature of supply chain risk, where the LLM itself acts as the interpreter of potentially malicious logic. This raises a fundamental challenge: how to identify and constrain harmful behavior when it is represented as semantically meaningful but syntactically benign text.

An open question is what forms of analysis—semantic, behavioral, or provenance-based—are sufficient to detect such attacks, and how these approaches can scale with the rapid growth of open agent ecosystems.

\subsection{Ecosystem Research}

\textbf{RQ5: What governance models can sustain open-source AI agent ecosystems while mitigating systemic supply chain risks at scale?}

As discussed in Section~\ref{sec:ecosystem}, OpenClaw's plugin-based ecosystem enabled rapid growth but also expanded the attack surface, while its permissive licensing and lack of built-in governance mechanisms provided no structural safeguards against malicious contributions. The ClawHavoc campaign illustrates how such a largely permissionless contribution model can lead to large-scale compromise in agent skill marketplaces.

In contrast, tightly controlled ecosystems such as Apple's App Store significantly reduce supply chain risk through centralized review, but at the cost of reduced openness and developer flexibility. This contrast highlights a fundamental tension between accessibility and security in agent ecosystems.

While intermediate approaches---such as reputation systems, staged trust models, or automated screening---have been explored in other software ecosystems, their applicability and effectiveness in AI agent ecosystems remain unclear. An open question is how governance structures can be designed and evaluated to balance ecosystem growth, developer participation, and systemic security under adversarial conditions.

\textbf{RQ6: How do the architectural choices of single-agent (OpenClaw) versus multi-agent (CrewAI/AutoGen) systems affect fault isolation, security boundary enforcement, and emergent behavior at deployment scale?}

As discussed in Section~\ref{sec:ecosystem}, the distinction between single-agent and multi-agent architectures is one of the most fundamental design choices in the agent framework landscape, yet their comparative security and reliability properties remain poorly understood. The Moltbook phenomenon---emergent collaborative behaviors among multiple OpenClaw instances communicating through social networks~\cite{openclaw2026agent_runtime}\cite{manik2026moltbook}---suggests that multi-agent dynamics can arise even in systems not explicitly designed for them, with unclear implications for coordination, fault isolation, and security boundaries. In addition, recent evidence of cross-agent prompt injection propagation indicates that vulnerabilities may spread across interacting agents, but no formal threat model yet exists for such behaviors.

This raises an open question of how architectural choices across the single-agent/multi-agent spectrum shape blast radius of compromise, boundary enforcement, emergent coordination, and performance trade-offs in real-world deployments.

\subsection{Sociotechnical Research}

\textbf{RQ7: What factors explain the divergent regulatory responses to autonomous AI agents across jurisdictions, and how do existing AI governance frameworks apply to agentic systems?}

Recent developments suggest growing regulatory attention to autonomous AI agents across multiple jurisdictions, including emerging scrutiny of agent-level system access and data governance concerns. However, as discussed in Section~\ref{sec:adoption}, there is limited systematic analysis of how these emerging responses relate to existing AI governance frameworks.

While frameworks such as the EU AI Act and the NIST AI Risk Management Framework provide general approaches to risk classification and lifecycle governance, it remains unclear how well they capture the specific properties of agentic systems, including system-level access, persistent state, and autonomous multi-step action.

This raises an open question of how existing regulatory frameworks can be interpreted, adapted, or extended to address agentic AI, and what factors drive variation in regulatory responses across jurisdictions.

\textbf{RQ8: How can the security risk of autonomous AI agents be systematically characterized and quantified for enterprise adoption decisions, given the absence of standardized benchmarks and the rapid evolution of the threat landscape?}

As discussed in Section~\ref{sec:adoption}, there is a notable gap between community adoption and enterprise deployment, with highly popular agent platforms lacking corresponding enterprise uptake. This discrepancy is partly attributed to the absence of standardized methods for assessing security and operational risk in agent systems.

Existing efforts, such as the Personalized Agent Security Bench (PASB)~\cite{wang2026double_agent}, provide initial evaluation directions, but remain limited in scope and adoption. In practice, enterprise decision-making relies on heterogeneous sources, including vendor claims, security audits, and informal assessments~\cite{microsoft2026running_openclaw}\cite{certik2026openclaw_report}, resulting in inconsistent and difficult-to-compare evaluations.

These observations suggest an open question of how risk in agent systems can be defined, measured, and compared, particularly given their unique characteristics such as persistent state, tool access, natural language attack surfaces, and multi-step execution under dynamic conditions.

\subsection{Meta-Observations}

Three cross-cutting observations emerge from the research questions above:

\textbf{Security failures originate from missing boundaries across abstraction levels.} The challenges identified in RQ1--RQ4 collectively point to a common issue: the absence of clear separation between instruction, data, and execution across the agent stack. Whether at the architectural level (RQ1), human-in-the-loop control (RQ3), or ecosystem interfaces (RQ4--RQ5), vulnerabilities arise when control signals and operational effects are not cleanly aligned. This suggests that agent security is fundamentally a problem of boundary definition and enforcement, rather than isolated implementation flaws.

\textbf{Irreducible trade-offs drive ecosystem stratification.} The questions raised in RQ5 and RQ6 reflect a deeper tension between openness, flexibility, and security. Empirical evidence from the ecosystem analysis and derivative landscape shows that no single architecture simultaneously satisfies these objectives, leading to the emergence of specialized system tiers. Rather than convergence, the field is undergoing structured fragmentation, implying that future progress depends on interoperability and cross-system guarantees rather than universal design solutions.

\textbf{Adoption is constrained by unresolved sociotechnical alignment.} The issues highlighted in RQ7 and RQ8 indicate that technical capability alone is insufficient for widespread deployment. Gaps in governance models, regulatory interpretation, and risk quantification create barriers to enterprise and institutional adoption. This suggests that the long-term trajectory of agent systems will depend on aligning technical design with policy frameworks, organizational practices, and economic incentives.

\section{Conclusion}
\label{sec:conclusion}

This paper introduced ASTELD, a six-axis classification framework for autonomous AI agent platforms, and evaluated it through an eight-framework comparison and an in-depth OpenClaw case study. Rather than treating heterogeneous platforms as an undifferentiated list of capabilities, ASTELD represents each system through Architecture, Security, Tool integration, Execution, Level of autonomy and human control, and Deployment coordinates. Explicit category definitions, a dominant-configuration assignment rule, and range notation make the comparison operational while preserving important configuration variation.

The evaluation produced three principal results. First, ASTELD yielded distinct dominant profiles for all eight platforms in the study, demonstrating discrimination within the selected sample. Second, the mapped profiles exposed recurrent structural patterns: a security--accessibility diagonal, strong coupling between architecture and execution, and convergence in surface capabilities despite persistent architectural differentiation. Third, analysis of 50+ OpenClaw derivatives showed that modifications concentrate on Security, Execution, and Deployment---the axes on which the base platform faces its most visible constraints. This provides case-based evidence that ASTELD is not only descriptive but can organize explanations of ecosystem fragmentation.

OpenClaw supplies the empirical depth behind this evaluation. Its monolithic local-first architecture, reactive security posture, marketplace tool model, single-agent loop, and practical shift from guided to delegated autonomy produce a distinctive A1--S2--T2--E1--L2/L4--D1 profile. The architecture, six-category vulnerability taxonomy, institutional security assessments, adoption evidence, and derivative mapping show how those coordinates connect to observable outcomes: accessibility, viral growth, security debt, and specialized forks. The framework also identifies an important unoccupied region in the current sample: local-first personal deployment combined with enterprise-grade security (D1 + S4).

ASTELD should be interpreted as an empirical classification method rather than a universal score or causal model. Its present evaluation is limited to eight representative frameworks, dominant configurations, and evidence available at the time of analysis. Future work should test coding reliability across independent annotators, expand the platform sample, quantify distances between profiles, and evaluate whether ASTELD coordinates predict security incidents, adoption patterns, or framework-selection outcomes in larger datasets.

The central implication is that autonomous-agent platforms are not converging toward a single optimal architecture. Their design space is structured by persistent trade-offs among accessibility, security, expressiveness, autonomy, and deployment control. A common classification coordinate system makes those trade-offs explicit and provides a basis for cumulative comparison, system selection, and future empirical study.

As agent platforms continue to evolve, the practical question is not only which system has more capabilities, but which design coordinates produce those capabilities, risks, and governance requirements. ASTELD provides a concrete starting point for answering that question consistently.

\bibliography{references}

\end{document}